\documentclass[usenatbib,psfig]{mn2e}
\usepackage{epsfig}
\usepackage{amsmath}

\begin{document}

%%%%%%%%%%%%%%%%%%%%%%%%%%%%%%%%%%%%%%%%%%%%%%%%
                   
\title[SE-SNe in Disturbed Galaxies]{A Central Excess of Stripped-Envelope Supernovae within Disturbed Galaxies}  
\author[S. M. Habergham et al.]{S. M. Habergham$^1$ 
\thanks{E-mail: smh@astro.livjm.ac.uk (SMH)}, P. A. James$^1$, and
J. P. Anderson$^2$\\ $^1$Liverpool John Moores University, Birkenhead,
CH41 1LD, UK\\ $^2$Departamento de Astronom\'ia, Universidad de Chile,
Casilla 36-D, Santiago, Chile\\ }

\date{Accepted . Received ; in original form }

\pagerange{\pageref{firstpage}--\pageref{lastpage}} \pubyear{2011}

\maketitle

\label{firstpage}

\begin{abstract}

This paper presents an analysis of core-collapse supernova
distributions in isolated and interacting host galaxies, paying close
attention to the selection effects involved in conducting host galaxy
supernova studies. When taking into account all of the selection
effects within our host galaxy sample, we draw the following
conclusions:
\noindent
\begin{enumerate}
\item~ Within interacting, or `disturbed', systems there is a real,
  and statistically significant, increase in the fraction of
  stripped-envelope supernovae in the central regions. A discussion
  into what may cause this increased fraction, compared to the more
  common type IIP supernovae, and type II supernovae without
  sub-classifications, is presented. Selection effects are shown not
  to drive this result, and so we propose that this study provides
  direct evidence for a high-mass weighted initial mass function
  within the central regions of disturbed galaxies.\\
\item~ Within `undisturbed' spiral galaxies the radial distribution of
  type Ib and type Ic supernovae is statistically very different, with
  the latter showing a more centrally concentrated distribution.  This
  could be driven by metallicity gradients in these undisturbed
  galaxies, or radial variations in other properties (binarity or
  stellar rotation) driving envelope loss in progenitor stars. This
  result is not found in `disturbed' systems, where the distributions
  of type Ib and Ic supernovae are consistent.\\
\end{enumerate}
\end{abstract}


\section{Introduction}

The host galaxies of core-collapse supernovae (CCSNe) give crucial
information when trying to determine the progenitors of these
explosions, and have led to some important breakthroughs in the study
of CCSNe. These include the almost exclusive presence of CCSNe in
late-type galaxies \citep[e.g.][]{vand05} implying progenitors which
only occur in young stellar populations. More recently, host
properties played an important role in the discovery of a new
classification of supernovae termed `calcium-rich', and showing
spectroscopic characteristics of CCSNe yet originating in regions
devoid of star-formation \citep{pere10}.

Even more information can be gathered on the possible supernova
progenitors of the various subtypes of CCSNe, by looking at the
specific site of the SN explosion within the host galaxy. The CCSNe
group divides broadly into type II (SNII) and types Ib and Ic (SNIbc),
distinguished by the presence or lack of hydrogen in their spectra
respectively \citep{mink41}. The SNII subclass can be further divided
into IIP, IIL, IIn, IIb which will be discussed later; for a review of
supernova classifications see \citet{fili97}. Many studies have
investigated the association of these CCSNe subtypes with spiral arms
or HII regions \citep[e.g.][]{bart94,vanD92,vanD96,petr05}, with more
recent studies benefiting from increased statistics
\citep[e.g.][]{jame06,ande08,ande09,hako09}. These analyses indicated
that SNIbc are both more strongly associated with HII regions implying
higher mass progenitors, and more centrally concentrated within
galaxies, when compared to type II SNe, indicating a metallicity
dependence with observed SN-type. Both interpretations can be
explained theoretically by mass-loss via radiatively-driven stellar
winds whose effectiveness increases with progenitor mass
\citep{hege03}, or high metallicity environments
\citep[e.g.][]{puls96,hege03}.

In \citet{habe10}, henceforth HAJ10, the current authors carried out a
study on the distributions of CCSNe types in both `undisturbed' and
`disturbed' host galaxies. This study concluded that within the
disturbed systems an excess of SNIbc existed within the central
regions when compared to those of type II. A case study on Arp 299, a
so-called `supernova-factory' \citep{neff04,pere09}, also carried out
by the authors (\citealt{ande11} henceforth A11), found the same
unusual central excess of stripped-envelope supernovae (SE-SNe).
These results have ignited a discussion into the role selection
effects can play within host galaxy studies.

The analysis within this paper will concentrate on SE-SNe against
SNII+IIP (consistent with A11); where SNII are all type II supernovae
without sub-classifications, and SE-SNe include all type Ib, Ic, Ib/c,
IIb and IIL. Any type IIn supernovae (SNIIn) within the sample have
been omitted from the analysis and discussion due to uncertainties in
their progenitor stars, and often in their classification
\citep[e.g.][]{ande12,kiew12,kell11,vand00}.

The accumulation of subtypes into the SE-SNe group, rather than the
SNIbc and SNII groups of HAJ10, comes as a result of recent studies
into the progenitors of CCSNe and the emergence of a sequence of
progenitor envelope-stripping \citep{dess11}, or mass loss
\citep{hege03,crow07,geor09}. The sequence has type IIP retaining
their entire stellar envelope, followed by an increasing loss of
stellar envelope material from IIL, to IIb, Ib and Ic, which have lost
both their hydrogen and helium layers. The interpretation for this
increased envelope stripping has been debated. Theoretical studies
suggest that it could be due to an increase in the progenitor mass
\citep[e.g.][]{hege03}, the metallicity of the environment
\citep[e.g.][]{hege03}, the presence of a binary companion
\citep[e.g.][]{pods92}, or stellar rotation
\citep[e.g.][]{meyn03}. Observational evidence for any one particular
theory is difficult to determine, though there is evidence for both
metallicity (\citealt{ande09} hereafter AJ09, \citealt{modj11,lelo11}),
and progenitor mass \citep{ande08} playing a substantial role in the
stripping of stellar envelopes prior to explosion. The best
observational constraints would come from direct detections of CCSNe
progenitors on pre-explosion images. Although this technique has had
great success in a number of cases (e.g. \citealt{elia11,maun11}, see
\citealt{smar09} for a review on the topic), the need for very nearby
events severely limits the statistics available from these studies,
which thus far have only had success with the much more common type II
subclasses. In reality the cause is likely to be a combination of all
of these effects.

The nature of all CCSNe, having high-mass, short-lived progenitors,
provides a direct probe of recent or on-going star-formation within
their host galaxies. The differences between the different subtypes
also allows constraints to be placed on this star-formation. So
investigations into the host galaxies of CCSNe provide insights both
ways: into the possible progenitors of the supernovae themselves, and,
into the mode of star formation in the host.

Our previous papers, culminating in \citet{ande12}, have established a
probable mass sequence for CCSNe progenitors. Building on this the aim
of this paper is to test the robustness of the authors' previous study
into the variation of CCSNe distributions with environment (HAJ10). We
will present the data and host galaxy study sample in Section 2,
followed by a new analysis of the radial distribution of CCSNe within
their hosts in Section 3. A rigorous investigation into possible
selection effects present within the sample will be discussed in
Section 4. Section 5 will explore possible interpretations of the
results, namely the effects of SNe environment metallicity, and the
overall metallicity gradients of the hosts; the contribution of
stellar rotation and binarity to the possible progenitor population;
and the possibility of a modified initial mass function (IMF)
preferentially producing the most massive stars in the central regions
of `disturbed' host galaxies. Finally, in Section 6 we will summarise
the results and present our preferred interpretation and any possible
implications.

\section{Data Sample}

Our data are drawn from observations of host galaxies indicated by SN
discoveries contained within the Padova-Asiago SN
catalogue\footnote{http://web.pd.astro.it/supern} and
IAU\footnote{http://www.cfa.harvard.edu/iau/lists/Supernovae.html} SN
catalogues. Such catalogues inherently contain biases with some
resulting from initial SNe studies which targeted bright star-forming
galaxies, rather than the `blind' searches many groups are currently
pursuing (e.g. Palomar Transient Factory (PTF), \citealt{rau09,law09};
Pan-STARRS, \citealt{kais10}). Our analysis contains 218 host galaxies
containing 280 CCSNe of all subtypes, and although biases still
persist within this sample, until such sufficiently large samples are
accumulated through `blind' searches, this is the best analysis
possible. The large sample within this study also enables us to cover
more of the rarer subtypes of SNe, but given the selection effects
contained in the sample we do not claim that the relative numbers
found here are representative of the Universe as a whole, and the
conclusions drawn from the data are not dependent upon statistical
completeness.

These host galaxies have been observed over 15 years of observations
in both SNe-related and H$\alpha$ galaxy studies. This provides us
with a heterogeneous and randomly sampled selection of the to-date,
observed local (median host recession velocity $\sim$1918~kms$^{-1}$)
CCSNe.

All galaxies have H$\alpha$ narrow-band and {\em R-}broad band
imaging, and come from the following facilities: the MPG/ESO 2.2m
telescope at La Silla, Chile; and the Liverpool Telescope (LT),
Jacobus Kapteyn Telescope (JKT) and Isaac Newton Telescope (INT) all
on La Palma, the Canary Islands. The data were reduced in a standard
manner using routines in {\em Starlink} and IRAF\footnote{IRAF is
  distributed by the National Optical Astronomy Observatory, which is
  operated by the Association of Universities for Research in
  Astronomy (AURA) under cooperative agreement with the National
  Science Foundation.}. For more detailed information on the
observations and data reduction process refer to AJ08 and Anderson et
al. 2012 (submitted).

When analysing host galaxies by a characteristic such as disturbance,
classification errors can become problematic. In HAJ10 the
classification of galaxy disturbance was carried out by two
independent visual inspections of the host galaxy images. Although
more quantitative methods exist for classifying galaxy disturbance,
namely through the degree of asymmetry \citep{cons00,lotz04}, the
classification within this study assesses such a wide range of
parameters that it is better done via an eye-ball study than any one
quantitative single measure. The visual classification schemes
described in \citet{sura98} and \citet{veil02} have been widely used
\citep[e.g.][]{mira11,zamo11}, and in this paper we adopt a similar
technique, attributing galaxy characteristics to levels of
disturbance. Although the most disturbed galaxy systems often also
show signs of merger-triggered starbursts in the nuclei, there are
undoubtedly some galaxies which show signs of interaction without
displaying such energetic star-formation. For the reanalysis presented
in this paper the authors themselves reclassified each galaxy within
the sample as `disturbed' or `undisturbed' based on a number of
characteristics. An `extreme' sample of disturbed galaxies was chosen,
which contained hosts undergoing a major interaction with another
galaxy, or a which had large degree of asymmetry indicating a recent
tidal interaction with another system. A larger sample of `disturbed'
galaxies (including the `extreme' systems above) included those
displaying at least two signs of minor interactions. These signs were;
presence in a group with minor interaction, minor degrees of
asymmetry, shells, tails, irregular structure (small irregular systems
are not included in this analysis - this irregular structure is within
a larger star-forming galaxy), a double nucleus, extreme or irregular
dust patterns or a clumpy morphology.

A full list of classifications for each galaxy in our sample is
presented in Table 1. The full table is available online through the
supplementary material. \footnote{Where starbursts have been indicated
  these come from the presence of the galaxy within the GOALS sample
  \citep{armu09} as either a LIRG or ULIRG, or where the system is
  noted as a starburst on the NASA/IPAC Extragalactic Database (NED).}

\begin{table*}
  \centering
  \begin{minipage}{140mm}
    \caption{An excerpt of the galaxy classification information for
      our sample. The full table is available online.} 
    \begin{tabular}{lccccccccccccr}
      \hline
      \rotatebox{90}{Galaxy~~~~} 
      & \rotatebox{90}{Isolated} 
      & \rotatebox{90}{Group - no interaction} 
      & \rotatebox{90}{Group- minor interaction} 
      & \rotatebox{90}{Group - major interaction} 
      & \rotatebox{90}{Asymmetry - minor} 
      & \rotatebox{90}{Asymmetry - major} 
      & \rotatebox{90}{Shells} 
      & \rotatebox{90}{Tails} 
      & \rotatebox{90}{Irregular structure} 
      & \rotatebox{90}{Peculiar dust lanes} 
      & \rotatebox{90}{Clumpy morphology} 
      & \rotatebox{90}{Double} 
      & \rotatebox{90}{Starburst}\\
      \hline
      IC~391       & X &  &  &  &  & X &  &  &  &  &  & & \\
      MCG-02-14-03 &   &  & X &  & X &  &  & X &  &  &  &  & \\
      NGC1821      & X &  &  &  &  & X &  &  &  &  & X &  & \\
      NGC1832      &   & X &  &  &  & X &  &  &  &  &  &  & \\
      NGC1961      &   & & X & & & X &  &  & X &  &  &  & X\\
      IC~438       &   & X &  &  & X &  &  &  &  &  &  &  & \\
      IC2152       &   & X &  &  &  &  &  &  &  &  &  &  & \\
      NGC2207      &   &  &  & X &  &  &  &  &  &  &  & X & \\
      NGC2146      &   &  & X &  & X &  &  & X &  & X &  &  & X \\
      ESO121-G26   & X &  &  &  &  &  &  &  &  &  &  &  & \\
      \hline
    \end{tabular}
 \end{minipage}
\end{table*}

\section{Results}

We begin this section by testing whether the ratio of SE-SNe to SNII
differs with galaxy disturbance. The absolute rates of each SNe
sub-type would be subject to bias within our sample regarding host
galaxy selection.  However, between the different galaxy samples no
such bias is present and so the relative frequencies of each CCSNe
type give us useful information. Below we list the ratio of SE-SNe to
SNII in each sample, with the numbers of SNe in each group presented
in brackets. \\

\noindent
{\bf SE-SNe : SNII+IIP}
\begin{itemize}
\renewcommand{\labelitemi}{$\bullet$}
\item Undisturbed: 0.766 (59 : 77)  
\item Disturbed: 1.14 (65 : 57)
\item Extreme: 1.85 (26 : 14)
\end{itemize}

It is clear that the disturbed galaxies have a higher fraction of
SE-SNe, by more than a factor of two, when compared to undisturbed
galaxies.

We now present the analysis of the radial distribution of 258 CCSNe
(the 22 type IIn SNe within this sample will be the subject of a
future paper), in both `undisturbed' (136 SNe in 114 hosts) and
`disturbed' hosts (123 SNe in 89 hosts). The radial distribution is
defined in terms of the fraction of {\em R}-band light Fr({\em R})
which lies within the circle or ellipse which contains the SN. This
means that an Fr({\em R}) value of 0.0 corresponds to a supernova at
the central {\em R}-band peak of the galaxy emission, while a value of
1.0 implies an extreme outlying SN. A full description of this
analysis is given in AJ09.

\subsection{Radial Distribution Analysis}

In Figure 1 we present the CCSNe distribution in the isolated or
`undisturbed' galaxy sample, and in Fig. 2 in the `disturbed'
sample. The radial distributions are presented as histograms of the
Fr({\em R}) value in which the top panel presents all CCSNe, SNII
(type IIP and SNII without a sub-classification) in the middle panel
and all SE-SNe (Ib, Ic, Ibc, IIL and IIb) in the bottom panel. The
combination of SNIIP with SNII without sub-classifications is based on
the assumption that $\sim$70\% of these SNe would be SNIIP based on
Lick Observatory Supernova Search (LOSS) observed fractions
\citep{li11}. Indeed a Kolmogorov-Smirnov (KS) test of SNII+SNIIP
against SNIIP within the whole sample gives a probability that they
are drawn from the same parent distribution of $\sim$86\%.

\begin{figure}
\includegraphics[width=75mm,angle=270]{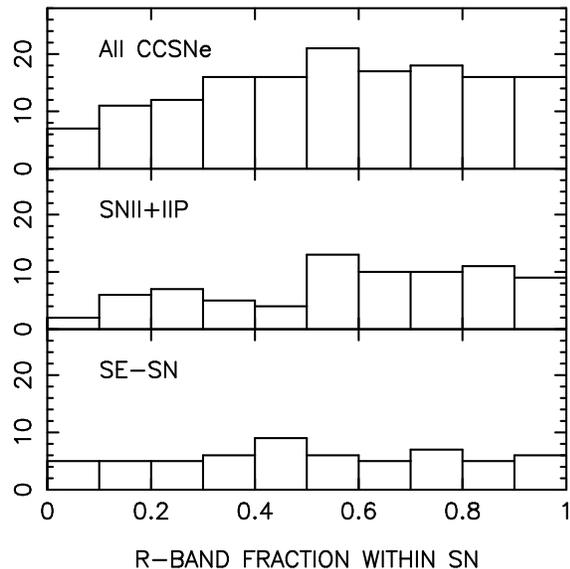}
\caption{Fractional {\em R}-band light distribution of CCSNe in
  `undisturbed' host galaxies.}
\label{fig:1}
\end{figure}

\begin{figure}
\includegraphics[width=75mm,angle=270]{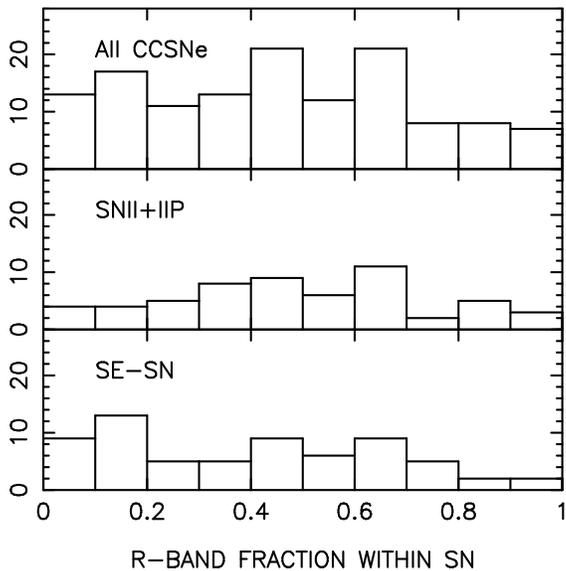}
\caption{Fractional {\em R}-band light distribution of CCSNe in
  `disturbed' host galaxies}
\label{fig:2}
\end{figure}

As discussed in Section 2, the classification criteria used to define
these samples mean that within the `disturbed' sample are galaxies in
all stages of interaction. This may be a galaxy within a group with
signs of minor asymmetry, to merging galaxy pairs. A subset of the
`extreme' galaxy sample can be seen in Fig. 3, and the results of the
radial analysis of the CCSNe in these systems in Fig. 4 with the same
panels as Figs. 1 and 2.

\begin{figure}
\includegraphics[width=75mm,angle=0]{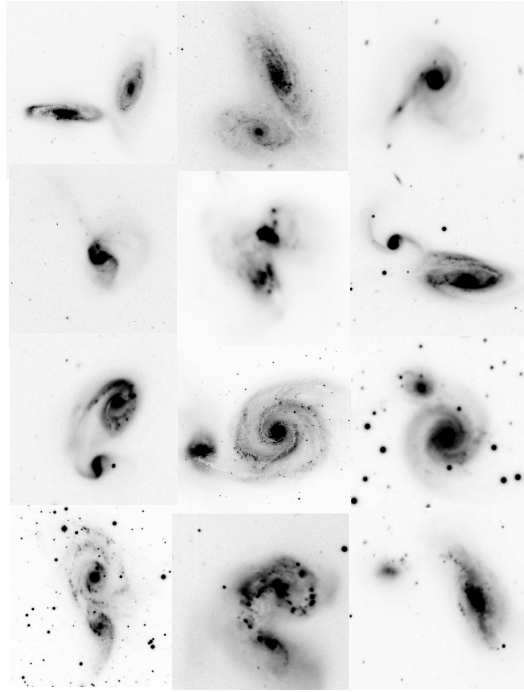}
\caption{Mosaic of 12 of the 28 host galaxies contained within our
  `extreme' sample.}
\label{fig:3}
\end{figure}

\begin{figure}
\includegraphics[width=75mm,angle=270]{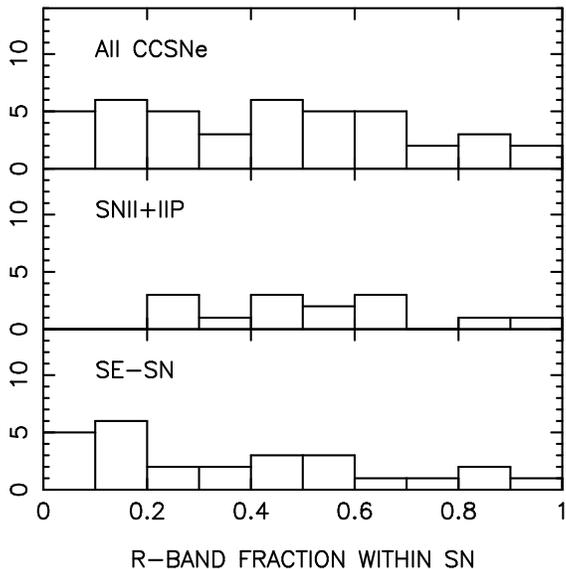}
\caption{Fractional {\em R}-band light distribution of CCSNe in the
  `extreme' host galaxy sub-sample}
\label{fig:4}
\end{figure}

%Any trends seen within the disturbed sample (Fig. 2) are therefore
%expected to become more exaggerated within the more `extreme' sample
%(Fig. 4), though the statistics diminish considerably.

The distributions of SNII+IIP and SE-SNe within each of the samples
appear markedly different. We present tests on the statistical
significance of these differences using a KS test. The KS test takes
two parameters to calculate the probability; the `distance' (the
largest difference between the distributions on a cumulative
distribution plot) between the distributions, and the number of events
within each distribution. Due to the decreased number of events in the
`extreme' sub-sample throughout the analysis of these results we will
present both the probability (P) and the distance (D) of the
distributions, in order to establish whether a reduced significance in
`P' is due to either a smaller sample size or a change in the
distribution. \\

%\noindent
%{\bf KS tests (II+IIP vs SE):}
%\begin{itemize}
%\renewcommand{\labelitemi}{$\bullet$}
%\item Undisturbed: P=0.106 D=0.2047 
%\item Disturbed: P=0.060 D=0.2337 
%\item Extreme: P=0.054 D=0.4231
%\end{itemize}
The results of the KS tests carried out in this study are presented in
Table 2, and will be discussed throughout the rest of this
section.\\ Within the `disturbed' galaxy sample there exists a
statistically significant (P$\sim$6~per cent) difference between the
distributions of SE- and IIP- supernovae; a result which becomes more
exaggerated within the `extreme' sample. This is a result that stands
from HAJ10 despite the addition of all SE-SNe into the same class.

\begin{table*}
  \centering
  \begin{minipage}{160mm}
  \caption{Kolmogorov-Smirnov test results for Section 3.}
  \begin{tabular}{lllllllll}
    \hline &\multicolumn{2}{c}{SN II+IIP vs
      SE-SNe}&\multicolumn{2}{c}{SN IIP vs SN
      Ibc}&\multicolumn{2}{c}{SN II+IIP vs SN
      Ibc}&\multicolumn{2}{c}{SN Ib vs SN Ic}\\ 
    \hline 
    & P & D & P & D & P & D & P & D \\ 
    \hline 
    Undisturbed & 0.106 & 0.2047 & 0.103 & 0.2566 & 0.086 & 0.2248 &
    0.004 & 0.5052 \\
    Disturbed & 0.060 & 0.2337 & 0.042 & 0.3171 & 0.014 & 0.2740 &
    0.208 & 0.2963 \\
    Extreme & 0.054 & 0.4231 & & & 0.035 & 0.4737 & & \\ 
    \hline
  \end{tabular}
  \end{minipage}
\end{table*}

In Figs. 5 and 6 we present the same data as in Figs. 1 and 2, but
without the intermediate SE-SNII classes (IIL and IIb), and without
type II SNe without sub-classifications, for `undisturbed' and
`disturbed' galaxies respectively. In both figures the top panel shows
the overall distribution of all CCSNe in the systems, the middle panel
the SNIIP and the bottom panel the SNIbc. The distributions shown
therefore represent the most extreme within our sample, where only SNe
which have lost their entire hydrogen envelope are plotted, and the
uncertainties in the SNII unclassified sample are removed.

Although the statistics are clearly reduced in these examples, the
central excess of SNIbc becomes even more apparent in the `disturbed'
galaxies. Despite the small statistics a Kolmogorov-Smirnov test
places the probability that SNIIP and SNIbc are drawn from the same
parent population within disturbed systems at $\sim$4~per cent. The
results of the KS tests are shown in Table 2.\\

\begin{figure}
\includegraphics[width=75mm,angle=270]{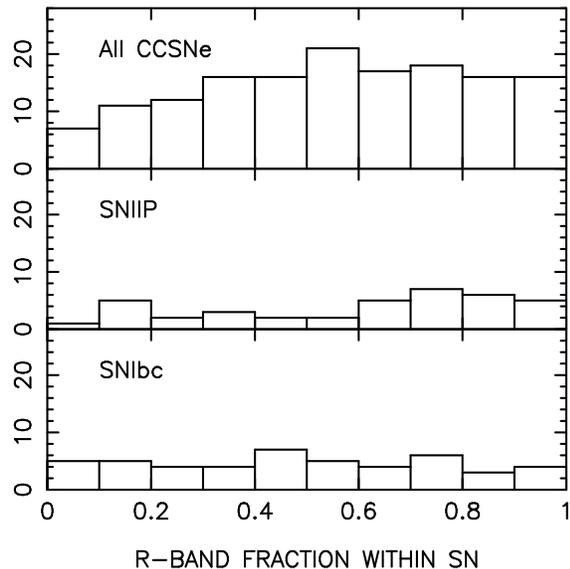}
\caption{Fractional {\em R}-band light distribution of type IIP and
  Ibc SNe in `undisturbed' hosts.}
\label{fig:5}
\end{figure}

\begin{figure}
\includegraphics[width=75mm,angle=270]{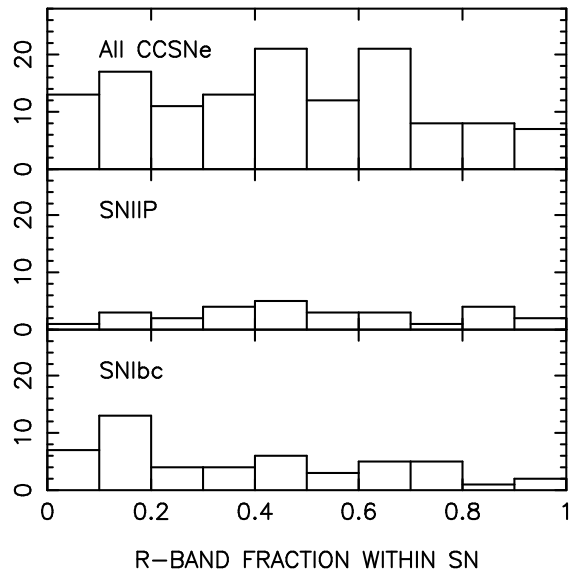}
\caption{Fractional {\em R}-band light distribution of type IIP and
  Ibc SNe in `disturbed' hosts.}
\label{fig:6}
\end{figure}

%\noindent
%{\bf KS tests (IIP vs Ibc):}
%\begin{itemize}
%\renewcommand{\labelitemi}{$\bullet$}
%\item Undisturbed: P=0.103 D=0.2566
%\item Disturbed: P=0.042 D=0.3171
%\end{itemize}

Combining the statistically consistent SNII population with the SNIIP
the results become more dramatic. Within the `disturbed' galaxies
there is only a $\sim$1~per cent probability that the SNII+SNIIP and
SNIbc populations have the same parent source. The reduced statistics
in the `extreme' sample raises this probability to $\sim$3.5~per cent,
although the large `D' value indicates that the distributions of these
populations are significantly different (Table 2).\\

%\noindent
%{\bf KS test (II+IIP vs Ibc):}
%\begin{itemize}
%\renewcommand{\labelitemi}{$\bullet$}
%\item Undisturbed: P=0.086 D=0.2248
%\item Disturbed: P=0.014 D=0.2740
%\item Extreme: P=0.035 D=0.4737
%\end{itemize}

\subsection{Distribution of SNIb and SNIc in undisturbed galaxies}

Where the statistics allowed we explored the distributions of all
CCSNe types, and in doing so another significant result has
emerged. Our samples are now large enough so that the SNIbc class can
be broken down into SNIb and SNIc individually. Within the
`undisturbed' sample we have 21 SNIb and 23 SNIc, and in the
`disturbed' sample, 21 SNIb and 27 SNIc. Figures 7 and 8 display the
histograms of the distributions of each population, where the top
panel is again the overall distribution of CCSNe, the middle panel
SNIb and the lower panel SNIc. 

The principal result here is that following a KS test, although the
distributions of SNIbc within the `disturbed' sample are completely
consistent (P$\sim$20~per cent), within the `undisturbed' sample
the two distributions are very different.\\

%\noindent
%{\bf KS test (Ib vs Ic):}
%\begin{itemize}
%\renewcommand{\labelitemi}{$\bullet$}
%\item Undisturbed: P=0.004 D=0.5052
%\item Disturbed: P=0.208 D=0.2963
%\end{itemize}

This places a probability that SNIb and SNIc, within `normal' spiral
galaxies, are drawn from the same parent population of only
$\sim$0.4~per cent (Table 2). Although the statistics here are small
this is an astonishing result, and not one which can be easily
explained in terms of selection effects.

The result seems to be driven in part by the lack of SNIb in the
central regions of these systems, and in part by the lack of SNIc in
the outer regions of `undisturbed' galaxies. The probability of
`missing' SNIc explosions in the outer regions of galaxies, where the
host extinction is low, seems unlikely. Correspondingly, the similar
peak absolute magnitudes of SNIb and SNIc \citep{li11}, mean that it
is unlikely SNIb are being missed in the central regions, where SNIc
are still observed.

The most likely explanation for the different distributions of SNIb
and SNIc in `normal' host galaxies is that the metallicity gradient
within the host plays a major role in evolving the progenitor star
prior to explosion. Envelope-stripping through radiatively driven,
metallicity-dependent, winds has been discussed in Section 1. We
believe the results presented here, showing that the distributions of
SNIb and SNIc within `undisturbed' galaxies are statistically very
different, provides direct evidence for this. 

One interesting aspect of our results is the apparent lack of SNIc in
the outer regions of our `disturbed' galaxies, where we would expect
some of the lowest metallicity regions. This echoes the findings of
\citet{arca10} who find a lack of SNIc in intrinsically low
metallicity dwarf galaxies.

The contribution of metallicity within `disturbed' host galaxies does
not appear to have as strong an effect as we will demonstrate in
Section 5.1.

\begin{figure}
\includegraphics[width=75mm,angle=270]{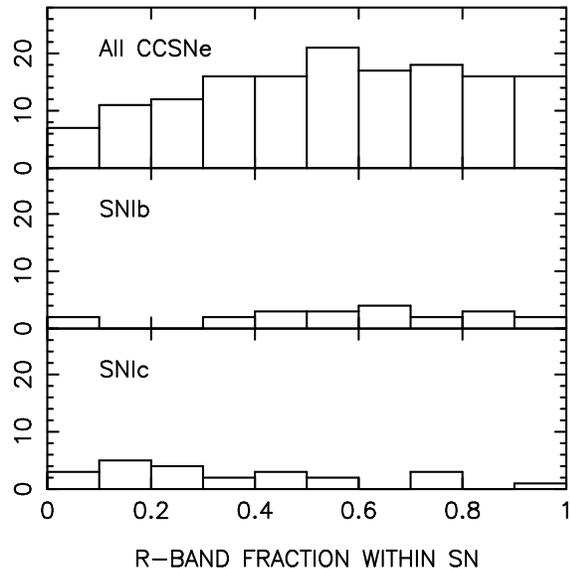}
\caption{Fractional {\em R}-band light distribution of Ib and Ic SNe
  in `undisturbed' host galaxies.}
\label{fig:6}
\end{figure}

\begin{figure}
\includegraphics[width=75mm,angle=270]{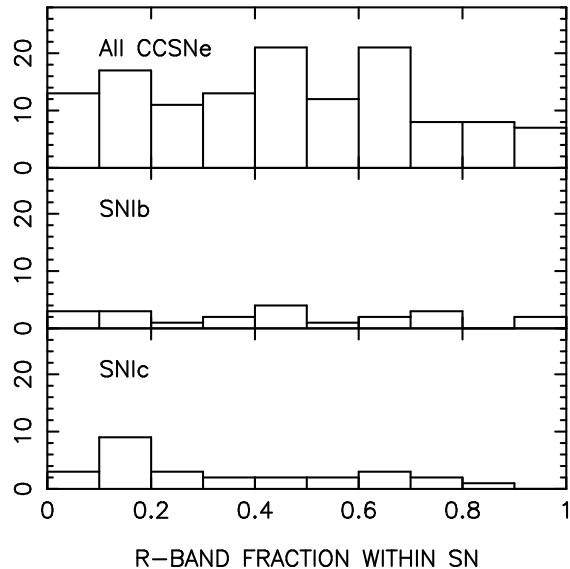}
\caption{Fractional {\em R}-band light distribution of Ib and Ic SNe
  in `disturbed' host galaxies}
\label{fig:7}
\end{figure}

\section{Selection Effects}

Within any host galaxy supernova study, selection effects and biases
can enter data from two possible sources: the supernova selection, and
the host galaxy selection. We shall explore the different effects
arising from each in this section.

\subsection{Supernovae}

A range of selection effects can contribute to biases within a CCSNe
study. The first of these is that different CCSNe types have different
magnitude ranges, both in peak magnitude and the duration of the light
curve, and so within any study, one is likely to lose intrinsically
fainter SNe subtypes, which are never discovered due to the brightness
of their host galaxy, or the extinction within it. It is often
accepted that SNIIP are most likely to be unaccounted for in large
volume galaxy studies \citep{past04,smart09}, due to their fainter
peak magnitudes during outburst. This effect could be exaggerated
within `disturbed' host systems due to the increased dust content, and
warped morphology. However, careful analysis of the SN luminosity
function from LOSS \citep{li11}, found no indication that SNIIP have
intrinsically fainter explosions. Although SNIIP have a faint tail
(not unlike SNIc), the absolute magnitudes of each CCSNe subclass have
enough overlap that missing SNIIP are unlikely to explain the results
presented in this paper.

The Shaw Effect \citep{shaw79}, that it is more difficult to detect
supernovae in the inner regions of distant galaxies, also affects all
supernovae searches and hence all host galaxy sample analysis.
Although the Shaw effect was much more applicable to photographic
plate searches, where the centres of galaxies were often over-exposed,
it is still a source of bias in today's SN searches.  Studies have
been carried out into the number of SNe estimated to be lost through
the Shaw effect. \citet{capp93} found that out to host galaxy
recession velocities of $\sim$6000 kms$^{-1}$ approximately 35~per
cent of {\em all} CCSNe types may be lost in photographic plate
searches in the inner regions of host galaxies. This compares to
$\sim$22~per cent of all SNe out to $\sim$4000 kms$^{-1}$ in some of
the first visual \citep{evan89} and CCD based automated searches
\citep{mull92}. Within this range of host galaxy redshifts there is no
preferential bias to any particular type of SNe though statistics at
this time were poor.  Studies into the Shaw effect on modern SNe
samples, assembled using modern CCDs and both targeted and `blind'
searches are lacking, and so for the purpose of this investigation the
results drawn from \citet{capp93} are the best estimates of bias
available in the literature.

To test further for selection biases, we analysed the distribution of
discovery magnitudes (taken from \citealt{lenn12} and the
Padova-Asiago SN catalogue \footnote{ http://web.pd.astro.it/supern/})
for our sample of SNIIP+II and SE-SNe. All SNe with available
discovery magnitudes were colour-corrected to the $R$-band, due to
this being closest to unfiltered light (a large proportion of the
discovery magnitudes). Our corrections were based upon available
CCSNe-subtype multi-band photometry on the SUSPECT database (the
online Supernova Spectrum Archive).

For this analysis we used the discovery magnitude of each SN rather
than a peak magnitude. In the context of exploring selection effects
within our sample, specifically whether we are missing `faint' SNe,
the discovery magnitude is the most important. Figure 9 shows the
absolute $R$-band magnitude at discovery of every CCSNe within our
sample, plotted against redshift. Here SE-SNe are represented by the
blue open circles, and SNII+IIP the red crosses. As expected, there is
a much larger range of discovery magnitudes within our sample at low
redshift. However, it is interesting to note that the brightest events
are found at all redshifts, and not just in the most distant galaxies.

\begin{figure}
\includegraphics[width=70mm,angle=270]{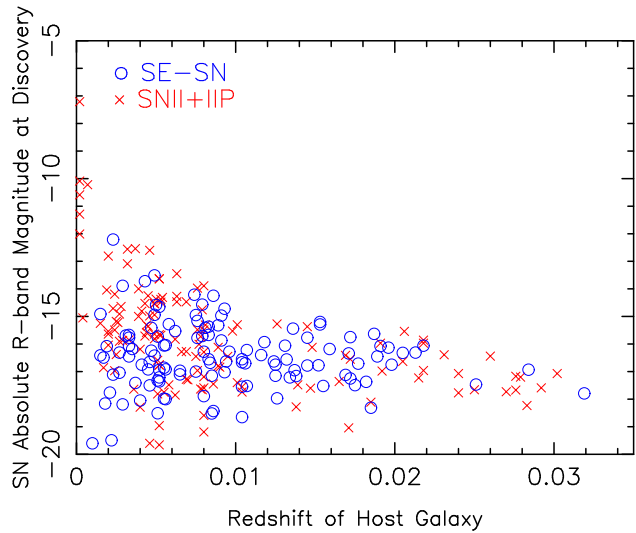}
\caption{Redshift distribution of all CCSNe in our sample against the
  absolute magnitude at discovery}
\label{fig:9}
\end{figure}

The importance of analysing the discovery magnitude of each SNe
against the Fr(R) value used in this paper is paramount in
establishing any possible selection effects. If we were to lose
`faint' SNe, we would do so in the central, dusty and high surface
brightness regions, defined in this paper as within $\sim$20~per cent
of the $R$-band light (Fr($R$)$<$0.2). Figures 10 and 11 display the
distribution of absolute and apparent discovery magnitudes,
respectively, of each SN in our sample against their Fr($R$)
value. Again, SE-SNe are represented by blue, open circles, and
SNII+IIP by red crosses.

Both Figs. 10 and 11 show no correlation between the discovery
magnitude (absolute or apparent) and the Fr($R$) value, for either the
SE-SNe or the SNII+IIP. Within the central regions we observe all
CCSNe types, over a similar range of magnitudes.\\

\begin{figure}
\includegraphics[width=70mm,angle=270]{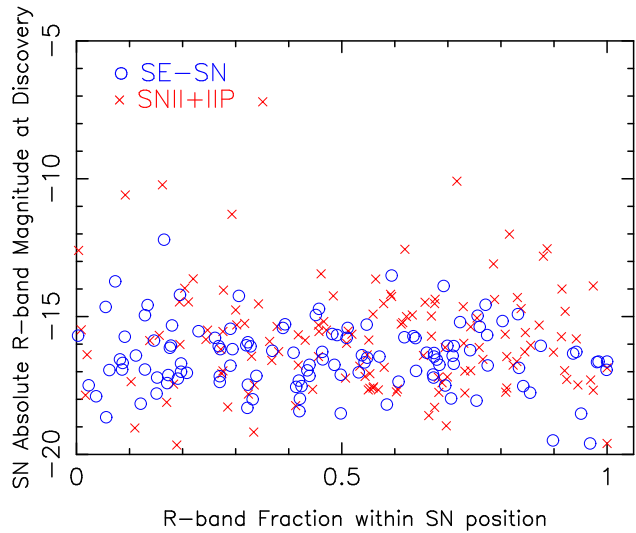}
\caption{Fractional {\em R}-band light distribution of each CCSN
  against its absolute magnitude at discovery.}
\label{fig:10}
\end{figure}

\begin{figure}
\includegraphics[width=70mm,angle=270]{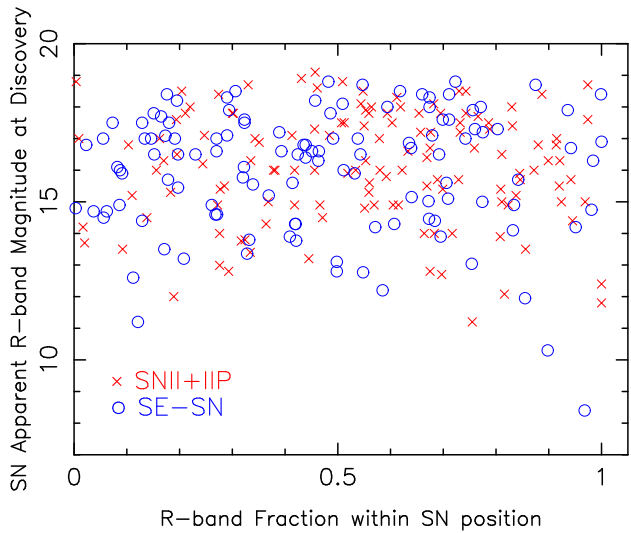}
\caption{Fractional {\em R}-band light distribution of each CCSN
  against its apparent magnitude at discovery.}
\label{fig:11}
\end{figure}

Further to this, it is important to analyse whether all of the central
SNe within our sample are present at the lowest redshifts. It may be
possible that the central SNe we detect are all in the most local
galaxies, and hence a selection effect could still affect the
result. In order to analyse this we have plotted the recession
velocity of each host galaxy and the Fr($R$) values of the SNe within
it. This is displayed in Fig. 12, where again SE-SNe are the blue open
circles and SNII+IIP are the red crosses. The dashed black lines
represent the median recession velocity of our sample and 6000
kms$^{-1}$, which was the limit of the HAJ10 study, and the limit to
which we would claim our results are reliable, due to the small number
statistics above this range.

Figure 12 again shows that there is no correlation between the Fr($R$)
value of the SNe and its host galaxy recession velocity. Certainly out
to $\sim$6000 kms$^{-1}$ there is no indication of losing any II+IIP
within the central regions, or at higher distances from the centre. A
KS test between low and middle redshift bins shows that the
distribution of SNII+IIP are completely consistent (P=0.990,
D=0.0814). The overall distributions of CCSNe within the 3 redshift
bins (z$<$median; median$>$z$>$6000 kms$^{-1}$; and z$>$6000
kms$^{-1}$) are all consistent with being drawn from the same parent
population.\\

We next test the radial distributions of CCSNe of all types within the
different redshift bins.\\

\noindent
{\bf KS test:}
\begin{itemize}
\renewcommand{\labelitemi}{$\bullet$}
\item low z bin~{\bf vs}~middle z bin:~P=0.947 D=0.0647
\item middle z bin~{\bf vs}~high z bin:~P=0.193 D=0.2353
\end{itemize}

This shows that these Fr($R$) distributions are statistically
completely consistent and that the main result presented in this
paper, of an excess of SE-SNe within the central regions of
`disturbed' hosts, is not due to a failure to detect SNIIP within
these regions.

Despite the small number of events in our sample above 6000
kms$^{-1}$, all of the histograms provided in Section 3, and all of
the KS tests given, include these events. Although the sample at this
redshift range is far from complete, the inclusion of such objects
within the results have no effect on the outcome, as the distribution
of events at all redshifts is essentially flat and therefore cannot
drive the results found.

In a very conservative test of possible distance dependent selection
effects, we now minimise these by analysing only those within 2000
kms$^{-1}$. This reduces the sample size in both `undisturbed' (67
CCSNe) and `disturbed' (56 CCSNe) samples. Even in this sample, a
central excess of SE-SNe is still visible, providing 11 of the 17
events within Fr($R$)=0.2 (compared to 3 out of 7 in the `undisturbed'
systems). However, at this level the result is no longer statistically
significant.

\begin{figure}
\includegraphics[width=70mm,angle=270]{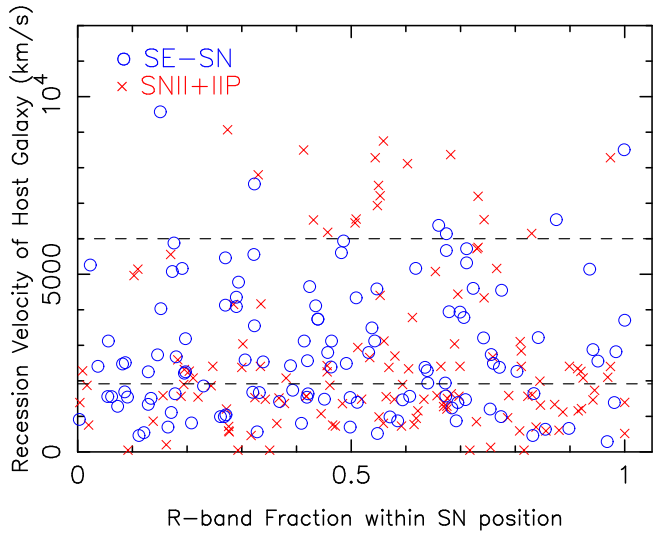}
\caption{Fractional {\em R}-band light distribution of each CCSN
  against the host galaxy recession velocity. The lower dashed line
  represents the median recession velocity for the sample, and the
  upper one the 6000km$^{-1}$ cut.}
\label{fig:12}
\end{figure}

Our radial analysis can also be carried out on the H$\alpha$ emission
within a galaxy, which traces star formation rather than the old
stellar light (traced by the $R$-band).  Here an Fr({\em H$\alpha$})
value of 0.0 means the supernova is closer to the central {\em R}-band
peak of the galaxy than any H$\alpha$ emission, and again a value of
1.0 indicates an outlying SN.

Figure 13 shows the histogram of Fr({\em H$\alpha$}) in our
`disturbed' galaxy sample. This figure shows that the central
distribution of SE-SNe is apparent, even with respect to the H$\alpha$
emission within the host galaxy. This suggests that it is not due to a
selection effect.

All of the investigations carried out in this section suggest that the
results found in Section 3 cannot be $driven$ by any selection
effect. We find in our data an $absolute$ excess of stripped-envelope
SNe within the central regions of `disturbed' galaxies, even with
respect to the star formation as traced by H$\alpha$.

\begin{figure}
\includegraphics[width=75mm,angle=270]{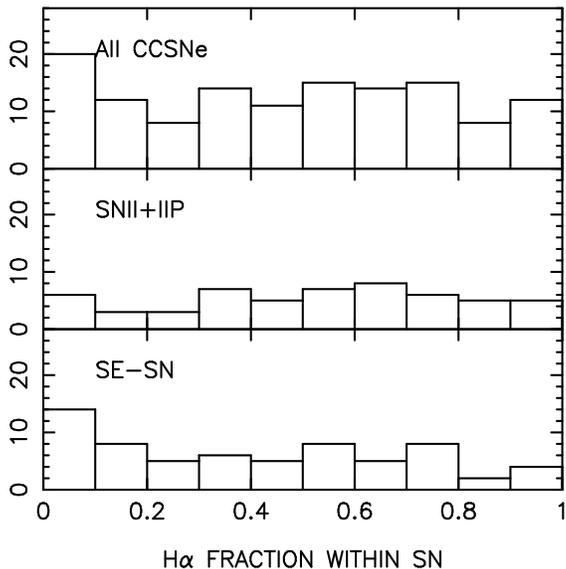}
\caption{Fractional {\em H$\alpha$} light distribution of CCSNe in
  `disturbed' hosts.}
\label{fig:13}
\end{figure}

\subsection{Host Galaxy Sample}

It is well documented that galaxy disturbance is linked to rapid
bursts of star formation \citep{lars78,jose84,kenn84,kenn87} which are
frequently concentrated in the central regions of the disturbed or
merging systems \citep{jose85,keel85}, probably linked to the central
gas concentrations seen in merger simulations
\citep{barn91,miho96}. Thus, a possible reason for the excess of
central stripped-envelope SNe in disturbed galaxies is that these
regions are dominated by starbursts of such extreme youth that they
have not yet started to produce SNIIP (under the assumption that CCSN
type is driven primarily by progenitor masses, and hence correlate
strongly with stellar lifetimes). 

In extremely young star formation regions, such as some observed in
Arp~299 \citep{alon00}, it is to be expected that current
supernovae should have high mass progenitors, regardless of the form
of the IMF.  However, this requires the galaxies to have been caught
at just the right age after an extremely short epoch of starburst
activity, which requires some fine tuning.

The degree of fine tuning required depends on the ranges of initial
mass of progenitors resulting in different types of supernovae, and in
particular the initial mass corresponding to the transition between
SNIIP and the stripped-envelope types.  This is a very uncertain
value, and depends on other factors, e.g. metallicity and
rotation. Estimates for the transitional mass between a type II and
SE-SNe for single-star progenitors include 25-40~M$_{\odot}$
\citep{hege03,eldr04} and $\sim$ 25~M$_{\odot}$ \citep{galy07}.
However, the rotating star models presented recently by \citet{ekst12}
and \citet{geor12} lower this value to between 15 and 20~M$_{\odot}$.
Observationally, the highest masses attributed to SNIIP progenitors
that have been detected in pre-explosion imaging are 15-18~M$_{\odot}$
for SN1999ev \citep{maun05}; 15~M$_{\odot}$ for SN2004dj
\citep{maiz04}; and 15~M$_{\odot}$ again for SN2004et \citep{li05},
although a recent study of the progenitor of SN2012aw \citep{fras12},
an unconfirmed SNIIP, found a higher mass range of 14-26~M$_{\odot}$.
From the statistics of all such progenitor detections found at the
time of publication, \citet{smar09} derived an upper limit to the
masses of SNIIP progenitors of 16.5~M$_{\odot}$.  Thus there is strong
evidence that stars at least as massive as 15~M$_{\odot}$ can produce
SNIIP, and conservatively taking this as the maximum mass, a burst of
star formation would start producing SNIIP after some 13~Myr,
according to the high-mass stellar lifetimes of \citet{meyn94}. If
instead the theoretically-preferred higher mass limits for this
transition are adopted, SNIIP could be occurring in as little as
5~Myr. While it is possible to catch one galaxy in a sufficiently
recent starburst phase that no SNIIP are yet occurring, for an
ensemble of galaxies, such as those presented here, and previously in
HAJ10, it seems extremely improbable that most could be caught at such
a critical phase of activity purely by chance.

One possibility is that our selection by morphological disturbance may
have resulted in systems with the very young starbursts required for
this timescale explanation. In an attempt to determine the
plausibility of this explanation, we used the results of
\citet{miho94,miho96}, who used Smoothed Particle Hydrodynamics
simulations to follow the behaviour of gas within minor and major
mergers of galaxies. These models contain a prescription for star
formation based on the original volume density formulation of the
Schmidt Law \citep{schm59}, enabling the calculation of the star
formation rate as a function of time. The overall timescale is defined
by a system of units that assumes the original galaxies, prior to
merging, to have had properties similar to the Milky Way. This is a
reasonable approximation for the bright interacting and merging
systems of interest in the present paper.

The details on the classification of galaxies within this study were
presented in Section 2 and Table 1. Figure 14 shows a subset of those
galaxies classified as `disturbed', which also hosted centrally-located
CCSNe. These have been matched up by eye with the stellar mass
distributions of the simulated mergers in \citet{miho96} and arranged
in an approximate chronological order, according to the corresponding
time within the \citet{miho96} simulations.  For the 6 systems shown,
the resulting merger ages are 260, 260, 689, 780, 845 and 910~Myr
respectively, from top left to bottom right.  The simulations of
\citet{miho96} also reveal the stages at which the central gas
densities, and hence the star formation activity, are predicted to be
strongly enhanced. This is the case for all but the first two frames
in Fig. 14, with the predicted starburst phase covering a simulated
time interval of over 200~Myr.

\begin{figure}
\includegraphics[width=90mm,angle=0]{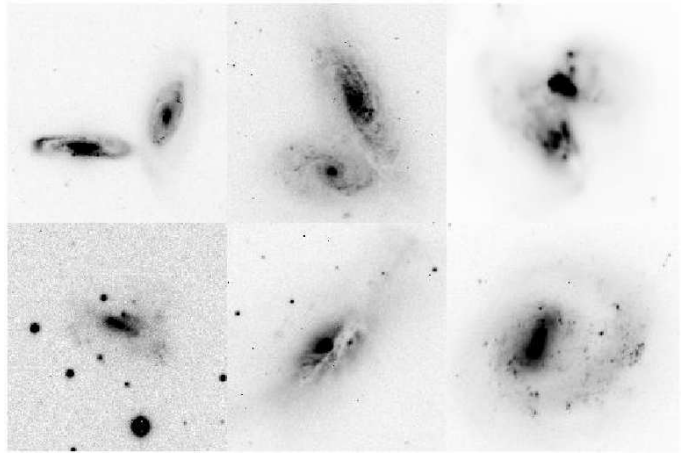}
\caption{Examples of the `disturbed' supernova-hosting galaxies from
  this study, arranged by the inferred stage of the
  merger/interaction.  Merger timescale increases from $\sim$260~Myr
  at top-left to $\sim$910~Myr at bottom-right.}
\label{fig:14}
\end{figure}

While the exact ages derived from this comparison of merger
morphologies with simulations are subject to substantial uncertainties
on a case by case basis, the overall timescales implied by the
differences in galaxy morphology for the observed sample cover a range
of hundreds of Myr, far in excess of the finely-tuned ages of a few
Myr required to select the `stripped-envelope dominated' phase at the
start of a starburst.  Even in the idealised case where a
merger-induced starburst can be represented by a narrow spike in the
central star formation rate, then for every one merger in the narrow
age window where only stripped-envelope SNe are being produced in
these regions, there should be several somewhat older starbursts
producing only SNII.  There is no observational evidence for the
latter phenomenon.

Apart from this fine-tuning argument based on the appearances of
galaxies and their dynamical timescales, there is further independent
evidence that observed starbursts have maintained their activity for
timescales easily long enough to be producing all types of CCSNe.
Studies of the energetics and spectroscopic properties of starbursts
generally infer ages of $\sim$40~Myr for the starburst phase
\citep{sand96,capu09}.  Of the most direct relevance for the present
argument is the detection of extended soft X-ray emission from the
regions surrounding starbursts \citep{fran03,iwas11}.  This emission
appears to be driven by CCSNe \citep{iwas11}, and the extent of the
emission implies that these SNe have been occurring for $\sim$10$^{8}$
years (K.~Iwasawa, private communication).  Again this is easily long
enough to have reached an equilibrium state in which all types of
CCSNe should be fairly represented.

Therefore we conclude here that the selection of young starbursts does
not drive the result presented here or in HAJ10.

\section{Interpretations}
Given that we have explored all possible sources of selection effect
and established that none can drive the central excess of SE-SNe in
disturbed galaxies, we now explore the possible interpretations of
this result. We look at each possible channel of progenitor
envelope-stripping, and how the environment within the central regions
of interacting galaxies may affect each channel.

\subsection{Metallicity}

As discussed briefly in the introduction, one key parameter thought to
lead to envelope-stripping, and hence the observation of a SNII or
SNIbc, is metallicity. Many SN host studies have observed a central
concentration of SNIbc when compared to SNII
\citep{bart92,vand97,tsve04,hako09,bois09,ande09}, as seen here, and
this has generally been assumed to be the result of metallicity
gradients present within the host galaxy. Within a `normal' star
forming galaxy, metallicity gradients have been observed, with the
central regions being more metal rich than the outer regions
\citep[e.g.][]{henr99}. Within this scenario central SNe are more
likely to lose their outer envelopes through metallicity-dependent
line driven winds \citep{puls96,kudr00,moki07}. All SNe host galaxy
studies to date have failed to distinguish between disturbed or
interacting systems and `normal' spirals within the sample (although
see \citealt{petr05} for a study of CCSNe with relation to
starbursts). Within merging galaxies any metallicity gradient
originally present is disrupted by the interaction and is smoothed out
within the system \citep{kewl10,rupk10,rich12}, through the infall of
pristine gas from the outer regions into the centre
\citep[e.g.][]{ramp05,hibb96,kewl06}, often fuelling a burst of star
formation \citep{barn96}. Furthermore, studies of the metallicity of
SNe sites have found little difference between regions that host SNII
and SNIbc, and a presence of both major groups of SNe at all
metallicities \citep{ande10}. This is supported by a study of NGC~2770
\citep{thon09} which found that the large number of SNIb within the
host were found in low metallicity regions, negating envelope
stripping via radiative winds. These studies suggest that the central
excess of SE-SNe in disturbed galaxies is not due to metallicity.

In order to study any metallicity gradient present within this sample,
we obtained spectroscopy for a random subset of the host
galaxies. Long-slit spectra for 23 of the host galaxies were obtained
using the Intermediate Dispersion Spectrograph (IDS) on the INT on La
Palma, in the Canary Islands. The slit was positioned to pass through
the nucleus of each galaxy, to give a positional reference, and angles
chosen to coincide with as many HII regions as possible. Metallicities
were determined using the line ratio diagnostics from \citet{pett04},
who use emission lines close together in wavelength to negate
extinction effects, which also addresses any atmospheric differential
refraction \citep{fili82} issues. In any case, observations were taken
where possible at the parallactic angle. It was not always possible to
obtain a spectrum at the SN-host HII region doing this and so where
possible (at low airmass) a second angle was observed to obtain
metallicity measurements for specific SNe sites. Standard reduction
procedures were employed using {\em Starlink} packages, and spectra
extracted from each observable HII region along the slit. Each
spectrum was then wavelength and flux calibrated. A subset of the
derived metallicity gradients is presented in Fig. 15. Each column in
Fig. 15 represents a different sample within the current study; the
first column displays host galaxies in the `undisturbed' sample, the
middle column `disturbed' galaxies, and the final column galaxies
within our `extreme' subset. Any galaxies containing known Active
Galactic Nuclei (AGN) had their central (bulge) metallicities
disregarded prior to determinations of the gradients. The median
metallicity gradients for each sub-sample are given below with the
number of galaxies in this sample following in brackets. 
\footnote{Throughout this section gradients are presented in units of
  $\frac{\Delta~log(O/H)}{\Delta~(R/R_{25})}$, where $R_{25}$ is the
  {\em B}-band isophotal radius at a surface brightness of 25 mag
  arcsec$^{-2}$.}

\noindent
\begin{itemize}
\renewcommand{\labelitemi}{$\bullet$}
\item Undisturbed: --0.5017(6) 
\item Disturbed: --0.4086(9) 
\item Extreme: --0.2447(7)
\end{itemize}

It should be noted here that the final figure changes dramatically (to
--0.5742) with the inclusion of NGC3627, which falls into our
`extreme' sample but shows an extremely, and probably unphysically,
steep metallicity gradient (--2.880) according to our measurements. It
is not known why this galaxy displays such a steep gradient. However,
it may also be interesting to note that were we to include the bulge
spectrum (initially excluded as the galaxy is classified as a LINER)
the gradient is reduced to --2.203, leading to an average within the
`extreme' sample of --0.4895.

\begin{figure*}
  \centering
  \begin{minipage}{160mm}
    \begin{tabular}{ccc}
      \resizebox{49mm}{!}{\includegraphics{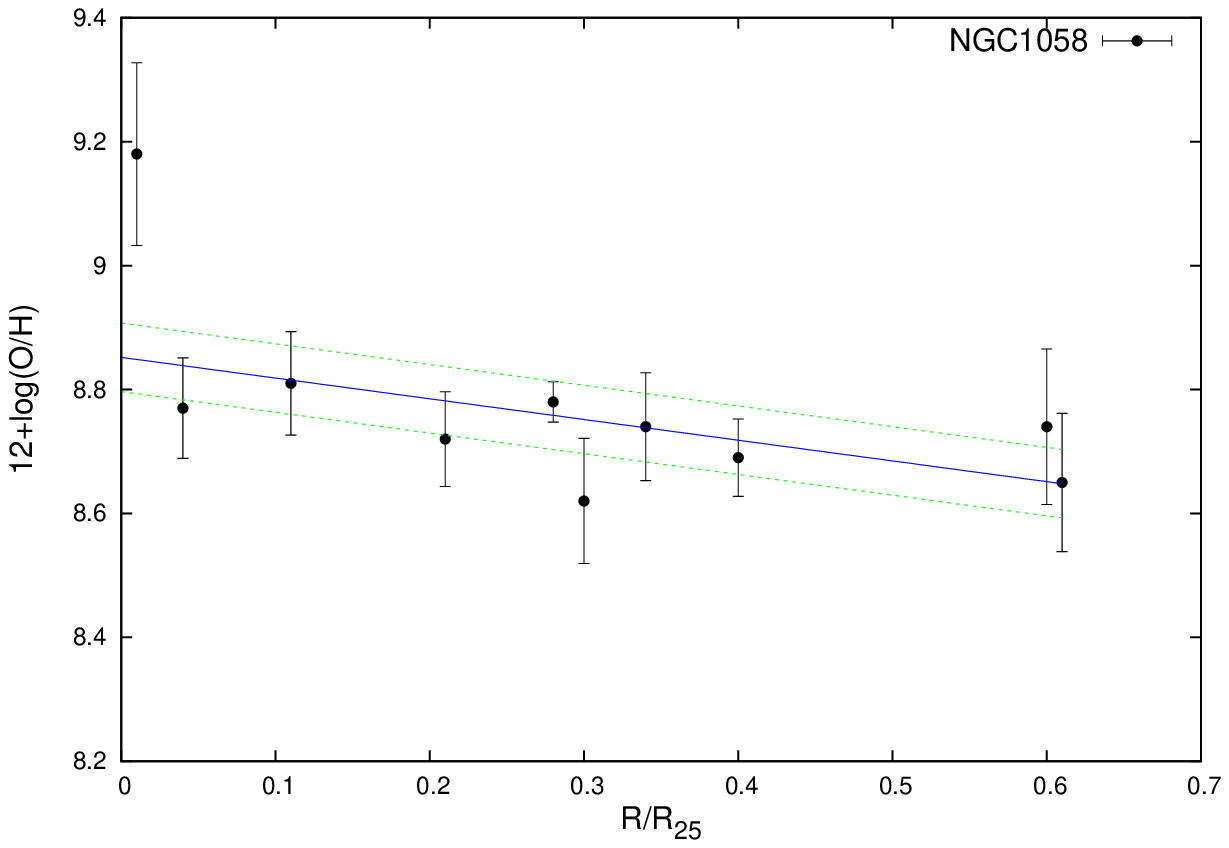}} &
      \resizebox{49mm}{!}{\includegraphics{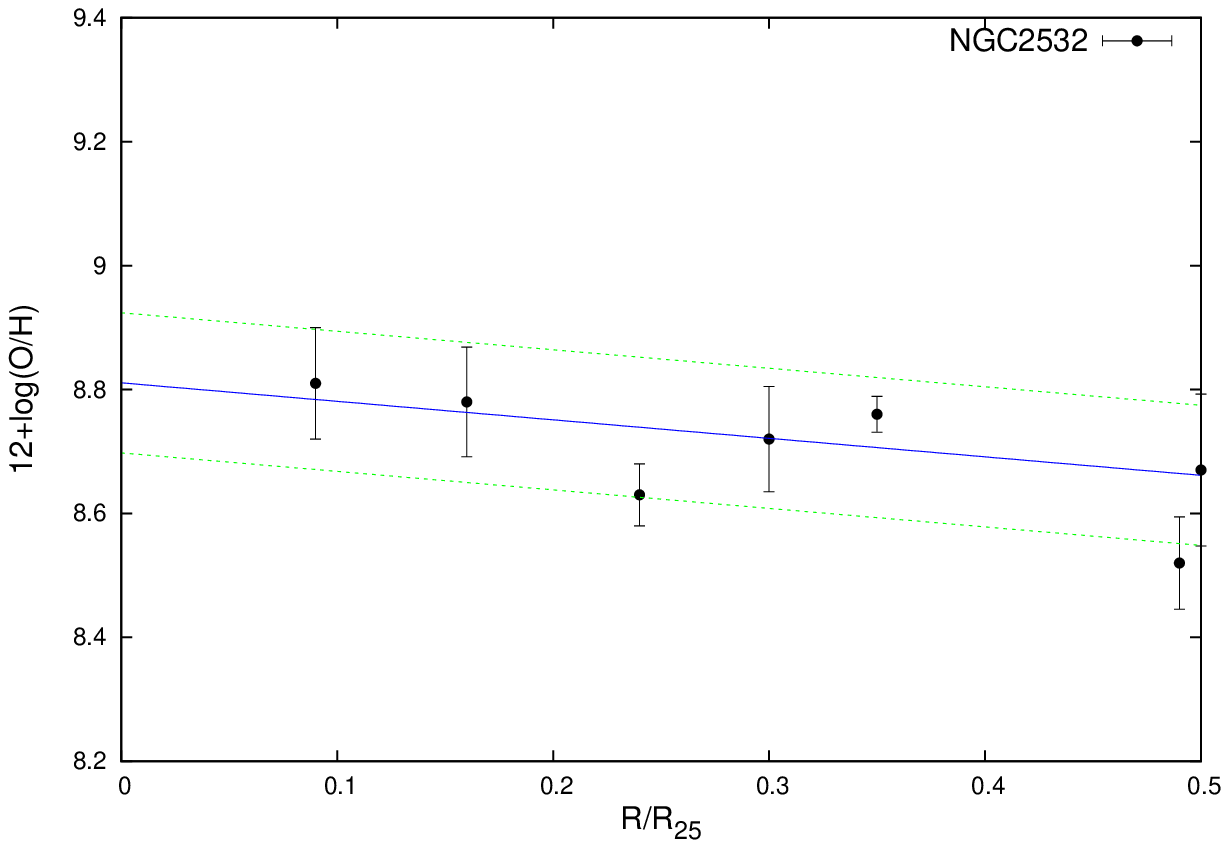}} &
      \resizebox{49mm}{!}{\includegraphics{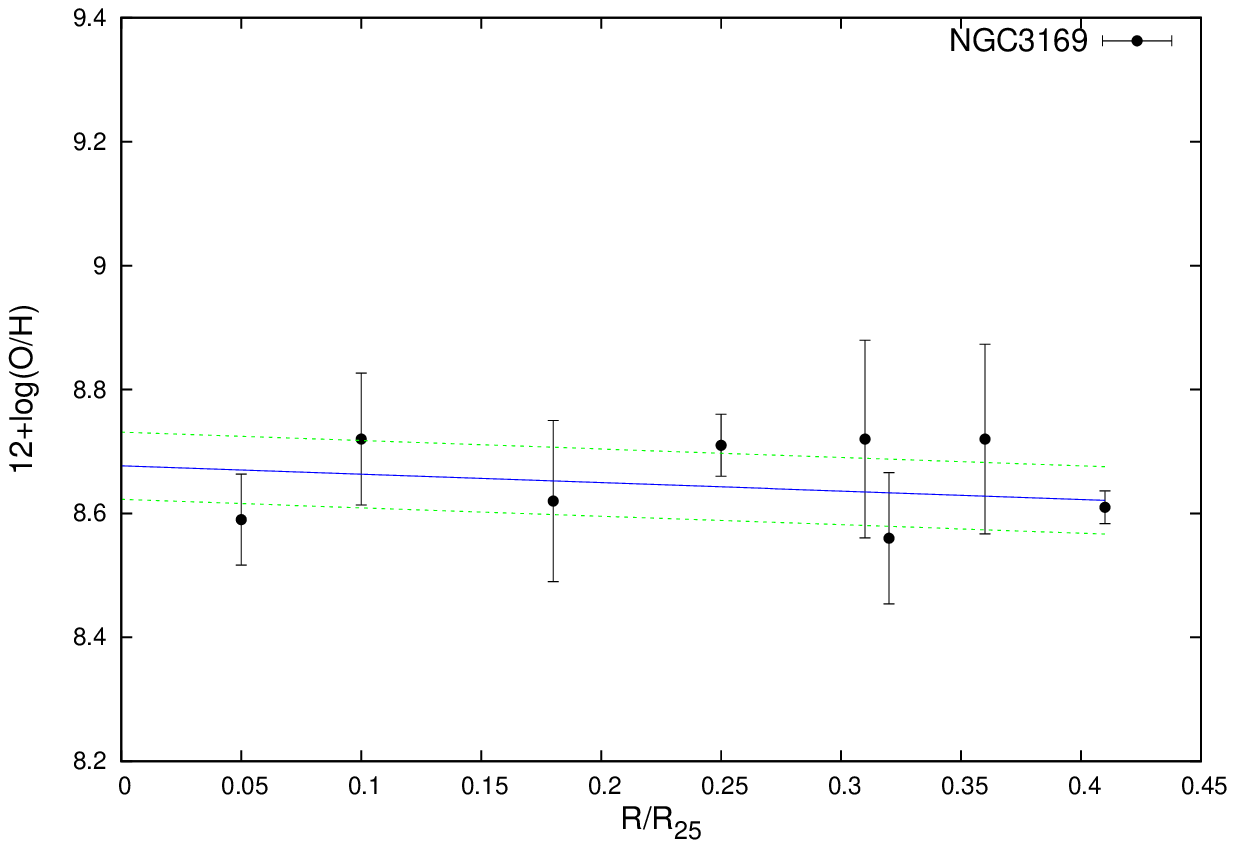}}\\
      \resizebox{49mm}{!}{\includegraphics{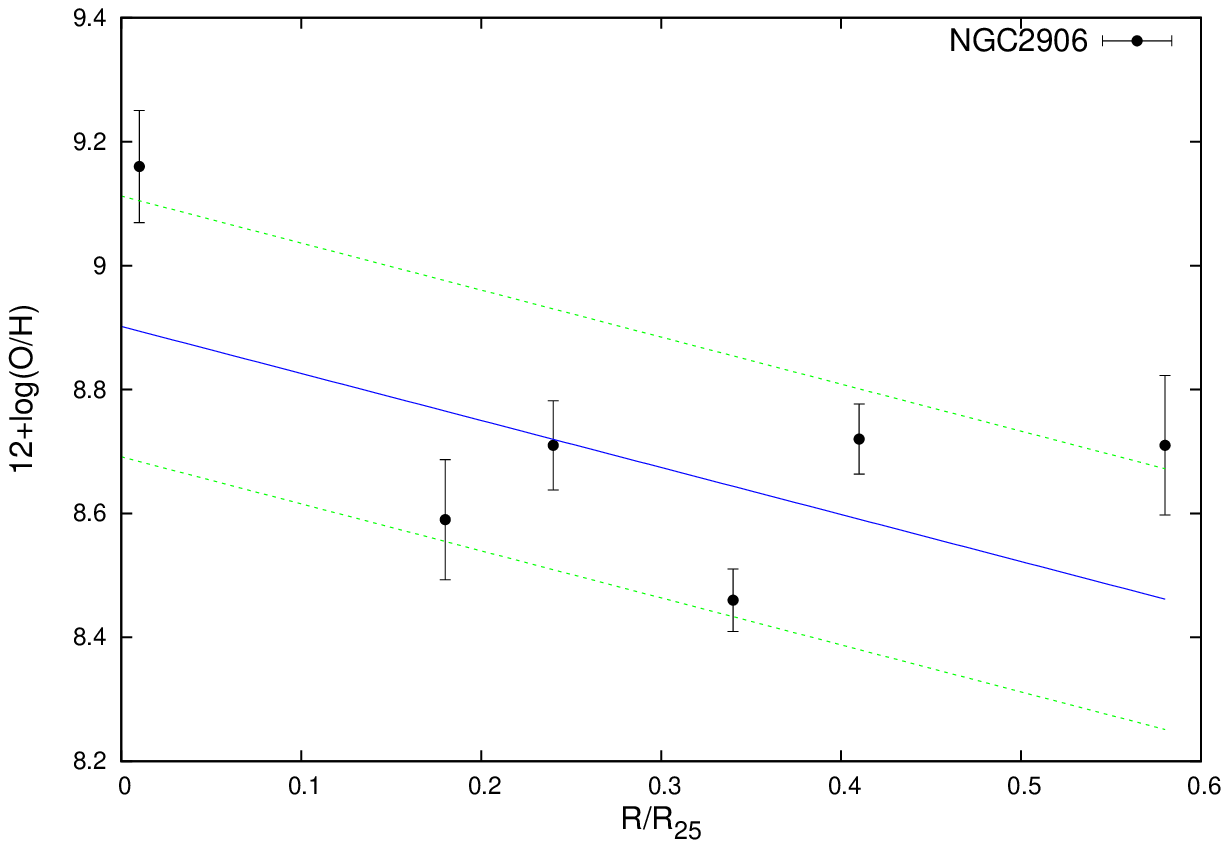}} &
      \resizebox{49mm}{!}{\includegraphics{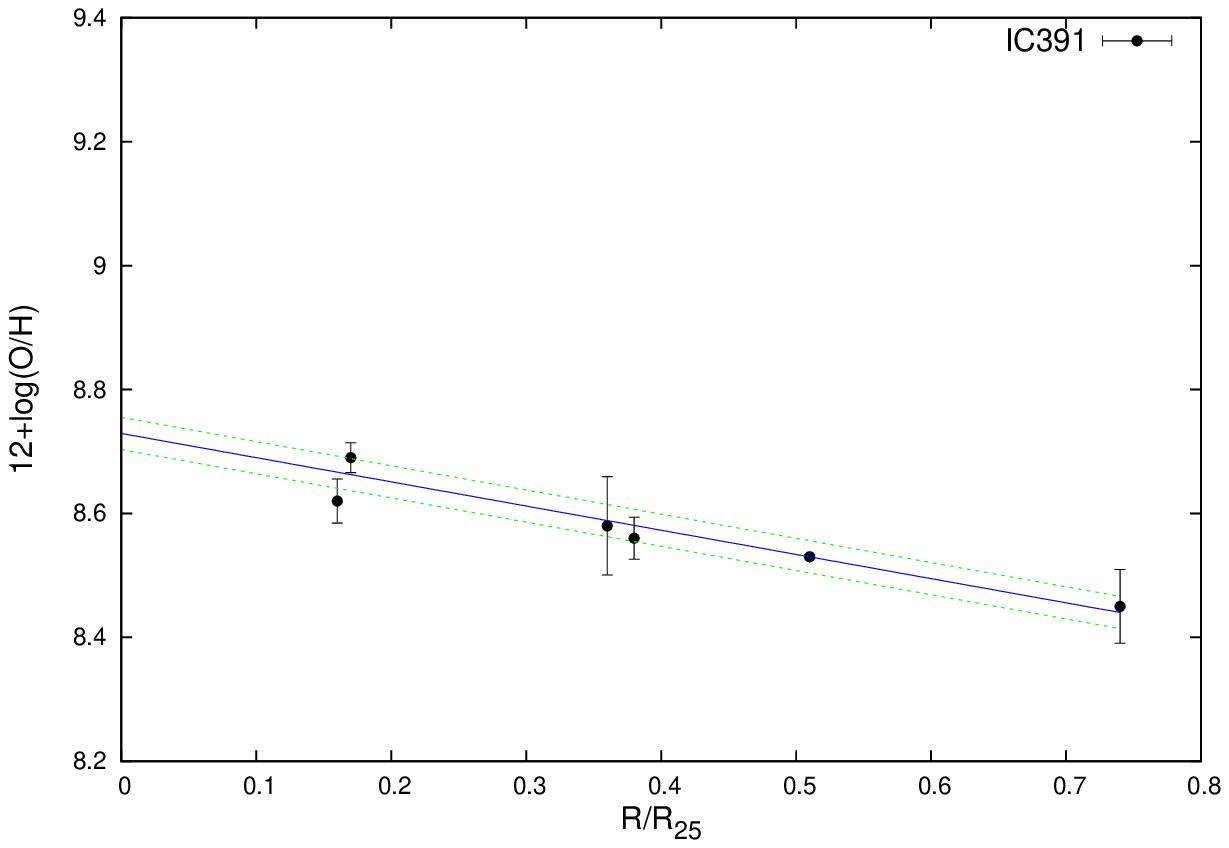}} &
      \resizebox{49mm}{!}{\includegraphics{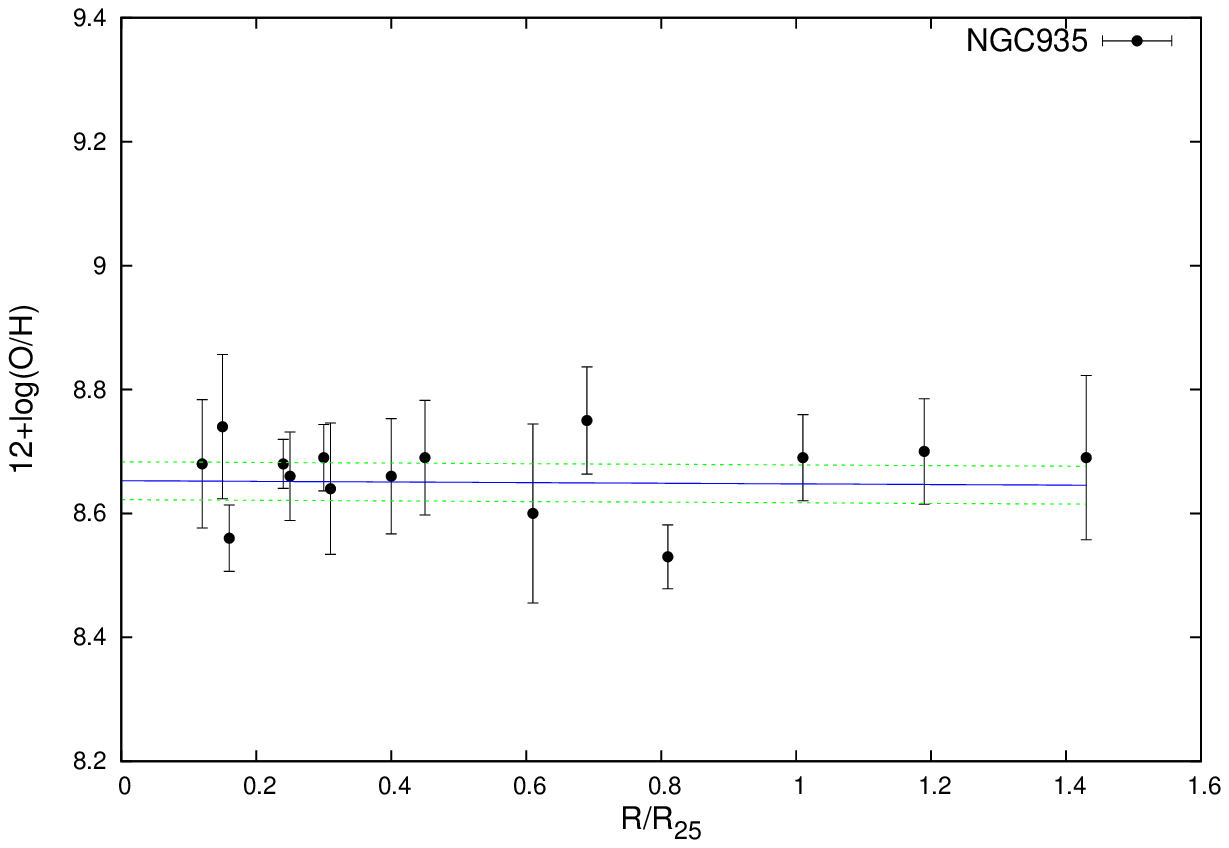}} \\
      \resizebox{49mm}{!}{\includegraphics{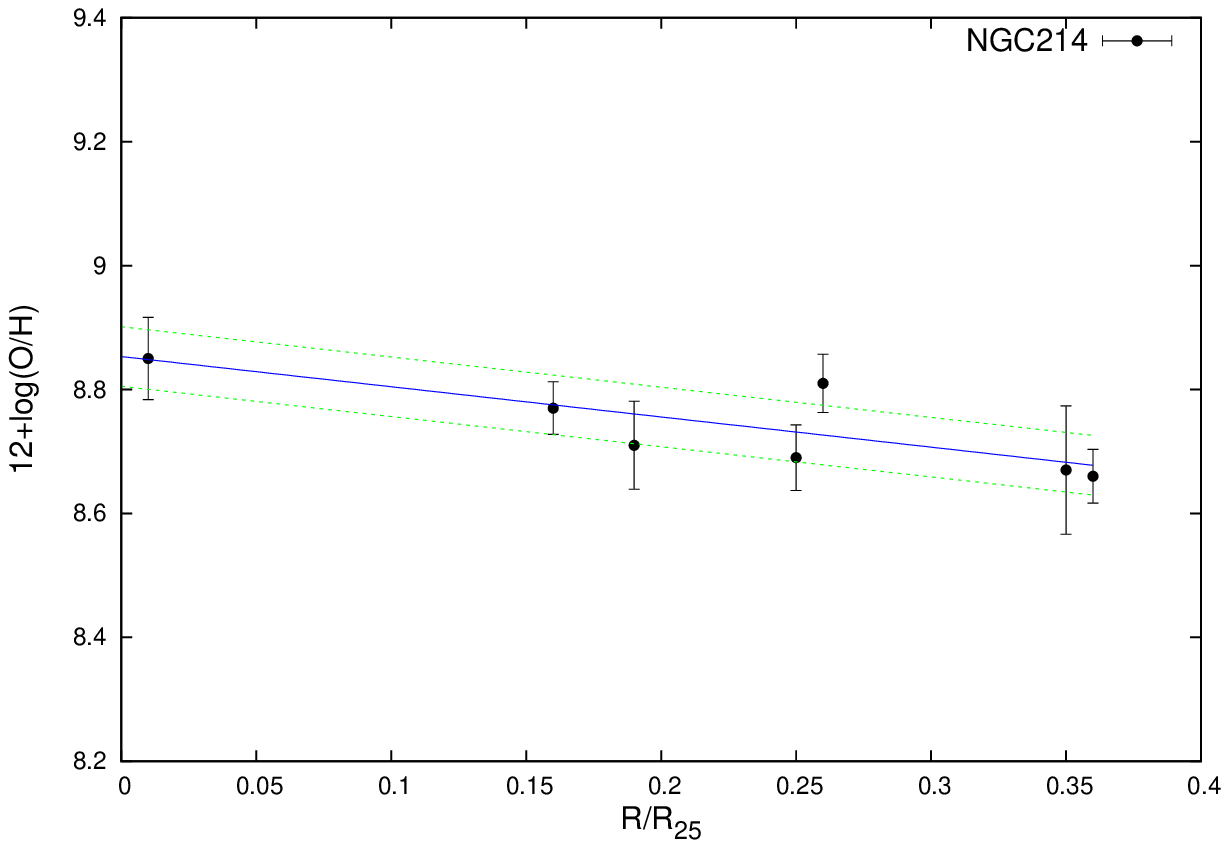}} &
      \resizebox{49mm}{!}{\includegraphics{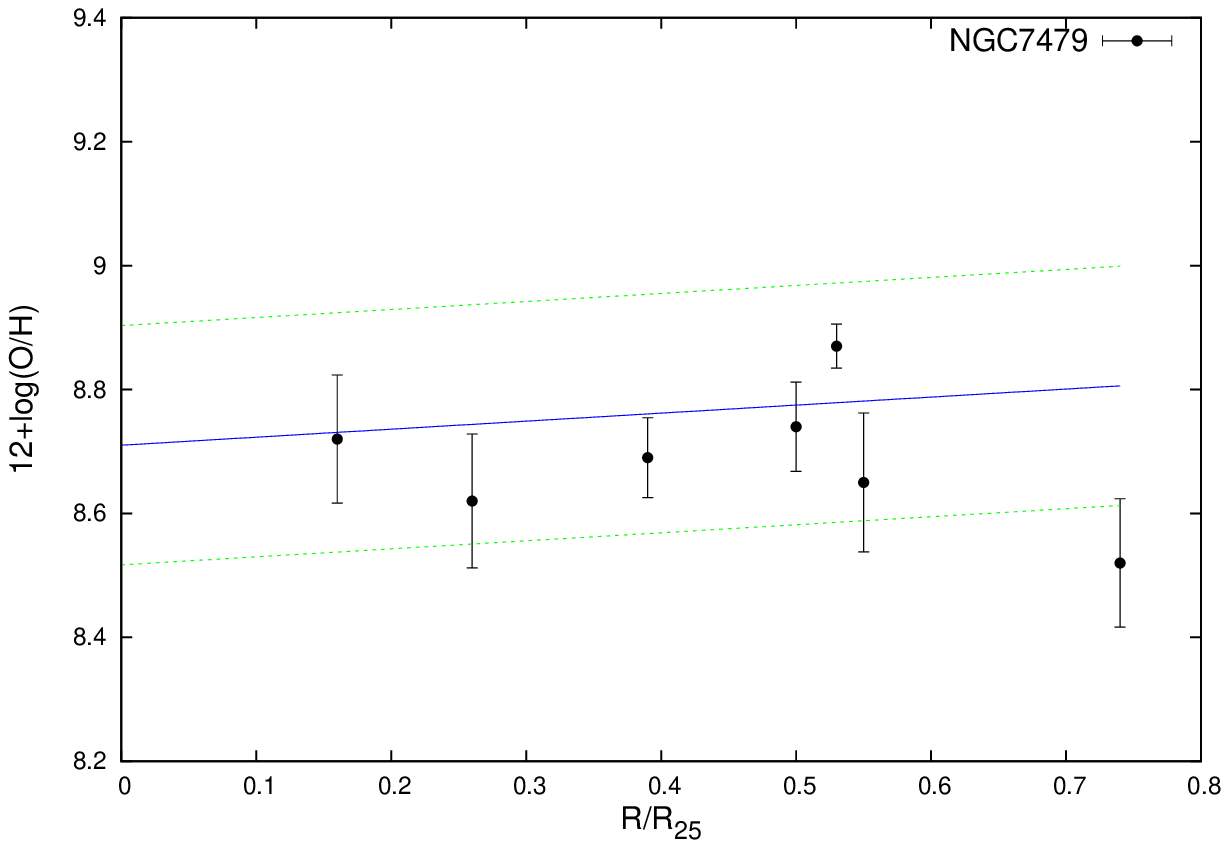}} &
      \resizebox{49mm}{!}{\includegraphics{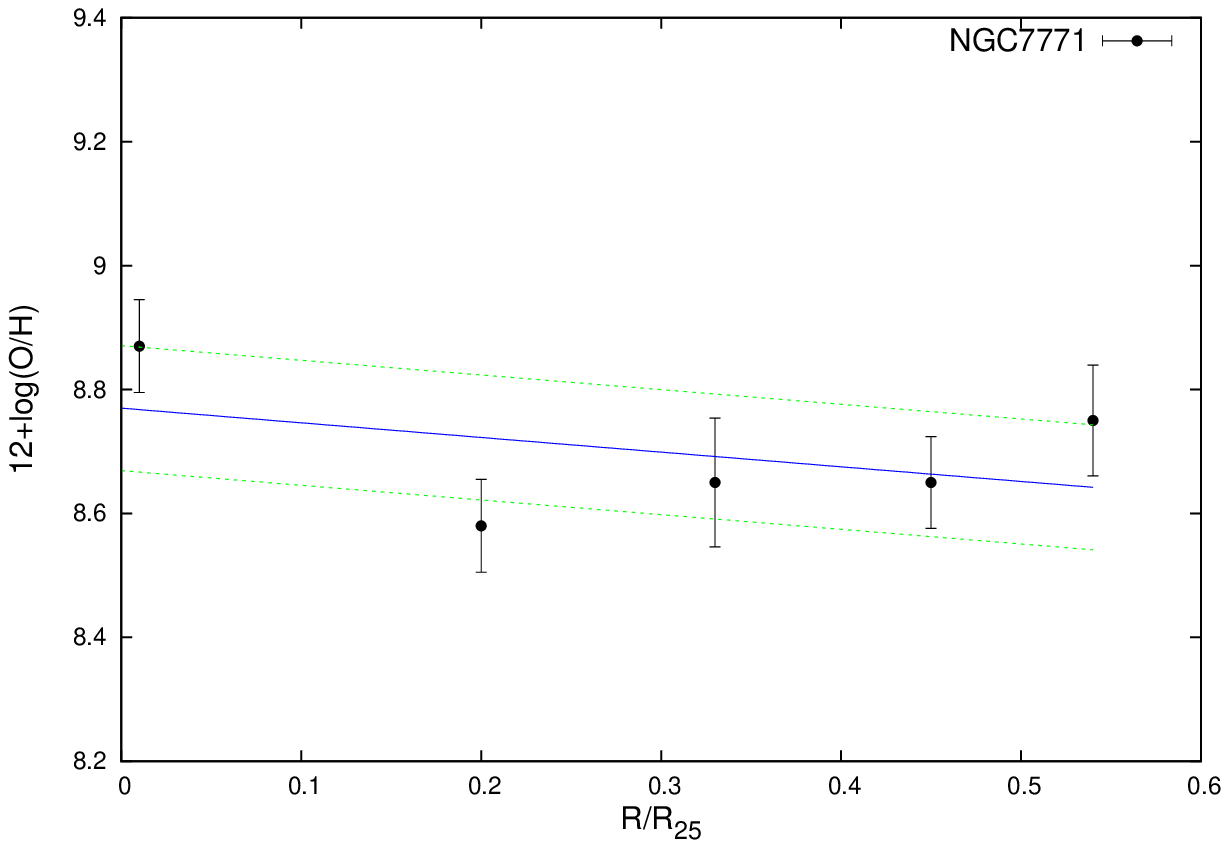}} \\
    \end{tabular}
    \caption{Metallicity gradients for a sample of the galaxies in INT
      IDS sample. `Undisturbed' galaxies are in the first column,
      `disturbed' systems in the middle column, and a selection of the
      `extreme' sub-sample in the right-hand column.}
    \label{fig:15}
  \end{minipage}
\end{figure*}

Our data overall offer support to the much more detailed study of
\citet{kewl10}, who find that the metallicity gradients become
shallower, or indeed disappear within disturbed or interacting
galaxies.

In accordance with \citet{ande10} it is also possible in some cases to
gather data on metallicity measurements at the SNe sites. Within our
`disturbed' sample there are 30 central CCSNe of all types (8 II+IIP
and 22 SE-SNe). Of these we possess host HII metallicity measurements
for 13 (from \citet{ande10} and the recent IDS data presented here),
all of which are classified as SNIb (7), SNIc (5) or SNIb/c (1).
Table 3 presents the average metallicity for each SNe type available
here compared to recent studies by \citet{ande10}, \citet{modj11} and
\citet{lelo11}. The table shows that our central SNe have
metallicities that are completely consistent with those presented in
other studies. The SNIc within the sample are entirely consistent with
all of the studies, whereas the SNIb average metallicity is more
consistent with \citet{lelo11}.

\begin{table*}
  \centering
  \begin{minipage}{160mm}
  \caption{Comparing the average metallicities for central SNIb, SNIc
    and SNIbc from this study, with the literature mean values of
    \citet{ande10,modj11,lelo11}.}
  \begin{tabular}{lccccccccc}
    \hline Study & SNIb metallicity & $\sigma$ & N & SNIc metallicity
    & $\sigma$ & N & SNIbc metallicity & $\sigma$ & N \\ 
    \hline 
    This paper & 8.581 & 0.041 & 7 & 8.637 & 0.067 & 5 & 8.617 & 0.035 & 13 \\ 
    \citet{ande10} & 8.616 & 0.040 & 10 & 8.626 & 0.039 & 14 & 8.635 & 0.026 & 27 \\
    \citet{modj11} & 8.49 & 0.012 & 13 & 8.66 & 0.010 & 14 & & & \\
    \citet{lelo11} & 8.52 & 0.05 & 14 & 8.60 & 0.08 & 5 & & &\\
    \hline
  \end{tabular}
  \end{minipage}
\end{table*}

The resulting conclusion from this discussion is that metallicity
cannot drive the excess of SE-SNe within the central regions of
interacting systems. Therefore alternate explanations must be pursued.

\subsection{Binary Progenitors}

The progenitors of CCSNe are widely discussed to arise from both
binary and rotating stars (see \citealt{eldr08,meyn05},
respectively). The relative contributions from each mechanism have
been discussed extensively in the literature
\citep[e.g.][]{pods92,geor12}. Even for models where binarity is
important, metallicity and mass may play an significant role
\citep{yoon10,eldr11,smit11}.

However, if we are to attempt to explain the central excess of SE-SNe
in disturbed galaxies in terms of binarity, we must analyse whether
binary fraction can be increased in such environments.

It is possible to analyse previous high density star forming regions,
by looking at present day globular clusters, which are thought to be
remnants of previous starburst episodes
\citep[e.g.][]{elme97,meur95,li04}. A recent study by \citet{krui12}
found that within merging galaxies the dynamical heating of star
clusters was an order of magnitude higher in interacting galaxies than
isolated ones, due to tidal shocks driven by the increased density.
This dynamical heating is sufficient to destroy star clusters at a
higher rate than new clusters are formed so that the total number of
stellar clusters in a merger remnant is $\sim$ 2-50~per cent of
the amount in the progenitor discs. However, within these shock heated
regions the massive clusters we would expect to host CCSNe, are more
likely to survive. Studies have shown that there is in fact an
anti-correlation between binary fractions of stars and the absolute
luminosity (mass) of the cluster \citep{milo08,soll07}.

Massive star clusters are known to have on average higher densities
and larger velocity dispersions \citep{djor93}. The fraction of
surviving binaries is thought to be dominated by binary ionisation and
evaporation \citep{soll08}, and so within high density regions
binaries are more likely to be disrupted through close encounters,
which are both more frequent and have higher mean kinetic energies
\citep{soll10}. However, during encounters between binary systems and
a massive single star the secondary star is usually ejected, therefore
in clusters with high collisional rates there will be an increase in
the number of equal mass binaries. \citet{soll08} found that this
increased fraction is not noticeable, with an exchange process of only
a few per cent over the entire cluster.

Therefore this suggests that the massive clusters which survive galaxy
interactions contain fewer binary systems. If these same arguments
apply in the high-mass stellar regime then the excess of central
SE-SNe cannot be explained in terms of an increased binary fraction.

\subsection{Modified Initial Mass Function}

In HAJ10, we characterised the modified IMF required to
accommodate our SN observations in terms of a (radically) modified
power-law index. However, there is to our knowledge no theoretical
motivation for an IMF that is flat or even rising with mass, and simply
changing the slope does not naturally explain the almost complete
absence of SNIIP we find.  Our results are more naturally accommodated
by suppressing low mass SF, up to some limit that might vary from
starburst to starburst, while leaving the form of the high-mass end of
the IMF unchanged.  To satisfy our observations, the requirement is
that this limit should be sufficiently high to prevent formation of
SNII progenitors in central regions of most (but not all) of our
`disturbed' galaxies.  Above this limit, the IMF can take the standard
form; this is unconstrained by our observations.  

This concurs with the theoretical models of \citet{kles07} who find
that in starburst galaxies the Jeans mass increases, which affects the
turnover of the IMF by pushing it to higher initial masses. This
`starburst' IMF is close to what is required to give the relative
numbers of SE- and IIP SNe found in our study. The IMF calculated for
the \citet{kles07} `starburst' simulation would lead to a strongly
suppressed number of SNII relative to SE-SNe, at least approximately
consistent with the results presented here, under the assumption that
progenitor mass differentiates these SN types.

Further discussion on this form of modified IMF and its implications
in the wider field of astronomy will be presented in a future paper
(James et al. in preparation).  

\section{Conclusions and Implications}

This paper has presented the results of a re-analysis of HAJ10,
including increased statistics and a deeper analysis of the
characteristics of disturbance. The results found can be summarised as
follows:

\noindent
\begin{itemize}
\renewcommand{\labelitemi}{$\bullet$}
\item We find a remarkable excess of SE-SNe within the central regions
  of `disturbed' galaxies. This confirms the main result from HAJ10,
  and is an absolute excess even with respect to the star formation as
  traced by H$\alpha$ emission.
\item The central excess is enhanced when we compare the SNIbc against
  SNIIP populations, as expected if a sequence of increased
  envelope-stripping exists from SNIIP-IIL-IIb-Ib-Ic.
\item We have found no preferential `loss' of SNIIP within the sample,
  both in terms of absolute and apparent discovery magnitudes, or with
  redshift within such a local sample of host galaxies. Therefore
  selection effects based around the SNe population studied here
  cannot drive the results found within this analysis.
\item The overall differences in galaxy morphology imply interaction
  timescales far in excess of those required to select the
  `stripped-envelope dominated' starburst phase within our
  sample. Therefore we conclude that there are no selection effects
  present within our host galaxies which could drive the central
  excess of SE-SNe in disturbed galaxies presented here.
\item Metallicity gradients within host galaxies cannot explain this
  result, as our disturbed sample are likely to have either a small
  gradient, or none at all. Furthermore, the metallicities of the
  local environments of the SNIbc within this sample are consistent
  with results published data.
\item The effects of stellar rotation and binarity have been discussed
  and no evidence has been found to indicate that either mechanism
  would be increased sufficiently within our host environments to
  explain the excess of SE-SNe found.
\item Our preferred explanation is that within the {\em central
  regions of disturbed galaxies a modified IMF exists which
  preferentially produces the most massive stars}, seen here in
  through the excess of SE-SNe.
\item The radial distributions of SNIb and SNIc within `undisturbed'
  galaxies are statistically very different (P=0.004). This could be
  driven by metallicity gradients in these undisturbed galaxies, or
  radial variations in other properties (binarity or stellar rotation)
  driving envelope loss in progenitor stars. This should be
  investigated in future studies.
\end{itemize}

The results of this investigation further constrain the progenitors of
different CCSNe subtypes, and highlight the need to gather as much
information as possible on the specific host galaxy environment, where
direct detections are impossible. We maintain that in different
environments the main mass-loss mechanism in place to produce
stripped-envelope supernovae can differ; most notably a metallicity
driven wind in `undisturbed' hosts and a modified IMF in the central
regions of `disturbed' host galaxies. Further discussion of the
implications of a modified IMF in the central regions of `disturbed'
host galaxies is to follow (James et al. in prep). 
  
\section*{Acknowledgments}

We wish to thank John Eldridge, Steve Smartt and Paul Crowther for
useful suggestions and feedback following our earlier papers in this
area. SMH would also like to thank Joanne Bibby for her invaluable
advice and knowledge on spectral data reduction. This research has
made use of the NASA/IPAC Extragalactic Database (NED) which is
operated by the Jet Propulsion Laboratory, California Institute of
Technology under contract with the National Aeronautics and Space
Administration. The Liverpool Telescope is operated on the island of
La Palma by Liverpool John Moores University in the Spanish
Observatorio del Roque de los Muchachos of the Instituto de
Astrof\'isica de Canarias with financial support from the UK Science
and Technology Facilities Council. The Isaac Newton Telescope is
operated on the island of La Palma by the Isaac Newton Group in the
Spanish Observatorio del Roque de los Muchachos of the Instituto de
Astrof\'isica de Canarias.  Also based on observations made with the
ESO 2.2~m telescope at the La Silla Observatory (programme ID
084.D.0195).  JPA acknowledges fellowship funding from FONDECYT,
project number 3110142, and partial support from by Iniciativa
Cientifica Milenio through the Millennium Center for Supernova Science
(P10-064-F).  PAJ and SMH acknowledge the UK Science and Technology
Facilities Council for research grant and research studentship
support, respectively.

%\bibliographystyle{mn2e}
%\bibliography{refs}

\begin{thebibliography}{}

\bibitem[\protect\citeauthoryear{Alonso-Herrero et al.}{2000}]{alon00} Alonso-Herrero A., Rieke G.~H., Rieke M.~J., Scoville N.~Z., 2000, ApJ, 532, 845

\bibitem[\protect\citeauthoryear{Anderson \& James}{2008}]{ande08} Anderson J.~P., James P.~A., 2008, MNRAS, 390, 1527

\bibitem[\protect\citeauthoryear{Anderson \& James}{2009}]{ande09} Anderson J.~P., James P.~A., 2009, MNRAS, 399, 559 
 
\bibitem[\protect\citeauthoryear{Anderson et al.}{2010}]{ande10} Anderson J.~P., Covarrubias R.~A., James P.~A., Hamuy M., Habergham S.~M., 2010, MNRAS, 407, 2660 

\bibitem[\protect\citeauthoryear{Anderson, Habergham, \& James}{2011}]{ande11} Anderson J.~P., Habergham S.~M., James P.~A., 2011, MNRAS, 416, 567

\bibitem[\protect\citeauthoryear{Anderson et al.}{2012}]{ande12} Anderson J.~P., Habergham S.~M., James P.~A., Hamuy M., 2012, MNRAS accepted, arXiv:1205.3802

\bibitem[\protect\citeauthoryear{Arcavi et al.}{2010}]{arca10} Arcavi I., Gal-Yam A., Kasliwal M.~M., Quimby R.~M., Ofek E.~O., Kulkarni S.~R., Nugent P.~E., Cenko S.~B., Bloom J.~S., Sullivan M., Howell D.~A., Poznanski D., Filippenko A.~V., Law N., Hook I., J{\"o}nsson J., Blake S., Cooke J., Dekany R., Rahmer G., Hale D., Smith R., Zolkower J., Velur V., Walters R., Henning J., Bui K., McKenna D., Jacobsen J., 2010, ApJ, 721, 777

\bibitem[\protect\citeauthoryear{Armus et al.}{2009}]{armu09} Armus L., Mazzarella J.~M., Evans A.~S., Surace J.~A., Sanders D.~B., Iwasawa K., Frayer D.~T., Howell J.~H., Chan B., Petric A., Vavilkin T., Kim D.~C., Haan S., Inami H., Murphy E.~J., Appleton P.~N., Barnes J.~E., Bothun G., Bridge C.~R., Charmandaris V., Jensen J.~B., Kewley L.~J., Lord S., Madore B.~F., Marshall J.~A., Melbourne J.~E., Rich J., Satyapal S., Schulz B., Spoon H.~W.~W., Sturm E., U, V., Veilleux S., Xu K.,  2009, PASP, 121, 559

%\bibitem[\protect\citeauthoryear{Barbon et al.}{2009}]{barb09} Barbon R., Buondi V., Cappellaro E., Turatto M., 2009, yCat, 1, 2024 

\bibitem[\protect\citeauthoryear{Barnes \& Hernquist}{1991}]{barn91} Barnes J.~E., Hernquist L.~E., 1991, ApJ, 370, L65 

\bibitem[\protect\citeauthoryear{Barnes \& Hernquist}{1996}]{barn96} Barnes J.~E., Hernquist L., 1996, ApJ, 471, 115 

\bibitem[\protect\citeauthoryear{Bartunov, Makarova, \& Tsvetkov}{1992}]{bart92} Bartunov O.~S., Makarova I.~N., Tsvetkov D.~I., 1992, A\&A, 264, 428 

\bibitem[\protect\citeauthoryear{Bartunov, Tsvetkov, \& Filimonova}{1994}]{bart94} Bartunov O.~S., Tsvetkov D.~Y., Filimonova I.~V., 1994, PASP, 106, 1276 

\bibitem[\protect\citeauthoryear{Boissier \& Prantzos}{2009}]{bois09} Boissier S., Prantzos N., 2009, A\&A, 503, 137 

\bibitem[\protect\citeauthoryear{Cappellaro et al.}{1993}]{capp93} Cappellaro E., Turatto M., Benetti S., Tsvetkov D.~Y., Bartunov O.~S., Makarova I.~N., 1993, A\&A, 273, 383

\bibitem[\protect\citeauthoryear{Caputi et al.}{2009}]{capu09} Caputi K.~I., et al., 2009, ApJ, 707, 1387 

\bibitem[\protect\citeauthoryear{Conselice, Bershady, \& Jangren}{2000}]{cons00} Conselice C.~J., Bershady M.~A., Jangren A., 2000, ApJ, 529, 886

\bibitem[\protect\citeauthoryear{Crowther}{2007}]{crow07} Crowther P.~A., 2007, ARA\&A, 45, 177

\bibitem[\protect\citeauthoryear{Dessart et al.}{2011}]{dess11} Dessart L., Hillier D.~J., Livne E., Yoon S.-C., Woosley S., Waldman R., Langer N., 2011, MNRAS, 414, 2985

\bibitem[\protect\citeauthoryear{Djorgovski \& Meylan}{1993}]{djor93} Djorgovski S., Meylan G., 1993, American Astronomical Society Meeting Abstracts, 182, 50.15

\bibitem[\protect\citeauthoryear{Ekstr{\"o}m et al.}{2012}]{ekst12} Ekstr{\"o}m S., Georgy C., Eggenberger P., Meynet G., Mowlavi N., Wyttenbach A., Granada A., Decressin T., Hirschi R., Frischknecht U., Charbonnel C., Maeder A., 2012, A\&A, 537, 146

\bibitem[\protect\citeauthoryear{Eldridge \& Tout}{2004}]{eldr04} Eldridge J.~J., Tout C.~A., 2004, MNRAS, 353, 87 

\bibitem[\protect\citeauthoryear{Eldridge, Izzard \& Tout}{2008}]{eldr08} Eldridge J.~J., Izzard R.~G., Tout C.~A., 2008, MNRAS, 384, 1109 

\bibitem[\protect\citeauthoryear{Eldridge, Langer \& Tout}{2011}]{eldr11} Eldridge J.~J., Langer N., Tout C.~A., 2011, MNRAS, 414, 3501

\bibitem[\protect\citeauthoryear{Elias-Rosa et al.}{2011}]{elia11} Elias-Rosa N., van Dyk S.~D., Li W., Silverman J.~M., Foley R.~J., Ganeshalingam M., Mauerhan J.~C., Kankare E., Jha S., Filippenko A.~V., Beckman J.~E., Berger E., Cuillandre J.~C., Smith N., 2011, ApJ, 742, 6

\bibitem[\protect\citeauthoryear{Elmegreen \& Efremov}{1997}]{elme97} Elmegreen B.~G., Efremov Y.~N., 1997, ApJ, 480, 235

\bibitem[\protect\citeauthoryear{Evans, van den Bergh \& McClure}{1989}]{evan89} Evans R., van den Bergh S., McClure R.~D., 1989, ApJ, 345, 752

\bibitem[\protect\citeauthoryear{Filippenko}{1982}]{fili82} Filippenko A.~V., 1982, PASP, 94, 715 

\bibitem[\protect\citeauthoryear{Filippenko}{1997}]{fili97} Filippenko A.~V., 1997, ARA\&A, 35, 309 

\bibitem[\protect\citeauthoryear{Franceschini et al.}{2003}]{fran03} Franceschini A., et al., 2003, MNRAS, 343, 1181 

\bibitem[\protect\citeauthoryear{Fraser et al.}{2012}]{fras12} Fraser M., Maund J.~R., Smartt S.~J., Botticella M.-T., Dall'Ora M., Inserra C., Tomasella L., Benetti S., Ciroi S., Eldridge J.~J., Ergon M., Kotak R., Mattila S., Ochner P., Pastorello A., Reilly E., Sollerman J., Stephens A., Taddia F., Valenti S., 2012, submitted, arXiv:1204.1523

\bibitem[\protect\citeauthoryear{Gal-Yam et al.}{2007}]{galy07} Gal-Yam A., et al., 2007, ApJ, 656, 372 

\bibitem[\protect\citeauthoryear{Georgy et al.}{2009}]{geor09} Georgy C., Meynet G., Walder R., Folini D., Maeder A., 2009, A\&A, 502, 611 

\bibitem[\protect\citeauthoryear{Georgy et al.}{2012}]{geor12} Georgy C., Ekstr{\"o}m S., Meynet G., Massey P., Levesque E.~M., Hirschi R., Eggenberger P., Maeder A., 2012, arXiv:1203.5243

\bibitem[\protect\citeauthoryear{Habergham, Anderson, \& James}{2010}]{habe10} Habergham S.~M., Anderson J.~P., James P.~A., 2010, ApJ, 717, 342 

\bibitem[\protect\citeauthoryear{Hakobyan et al.}{2009}]{hako09} Hakobyan A.~A., Mamon G.~A., Petrosian A.~R., Kunth D., Turatto M., 2009, A\&A, 508, 1259 

\bibitem[\protect\citeauthoryear{Heger et al.}{2003}]{hege03} Heger A., Fryer C.~L., Woosley S.~E., Langer N., Hartmann D.~H., 2003, ApJ, 591, 288 

\bibitem[\protect\citeauthoryear{Henry \& Worthey}{1999}]{henr99} Henry R.~B.~C., Worthey G., 1999, PASP, 111, 919 

\bibitem[\protect\citeauthoryear{Hibbard \& van Gorkom}{1996}]{hibb96} Hibbard J.~E., van Gorkom J.~H., 1996, AJ, 111, 655 

\bibitem[\protect\citeauthoryear{Iwasawa et al.}{2011}]{iwas11} Iwasawa K., Sanders D.~B., Teng S.~H., U V., Armus L., Evans A.~S., Howell J.~H., Komossa S., Mazzarella J.~M., Petric A.~O., Surace J.~A., Vavilkin T., Veilleux S., Trentham N., 2011, A\&A, 529, A106

\bibitem[\protect\citeauthoryear{James \& Anderson}{2006}]{jame06} James P.~A., Anderson J.~P., 2006, A\&A, 453, 57 

\bibitem[\protect\citeauthoryear{Joseph et al.}{1984}]{jose84} Joseph R.~D., Meikle W.~P.~S., Robertson N.~A., Wright G.~S., 1984, MNRAS, 209, 111 

\bibitem[\protect\citeauthoryear{Joseph \& Wright}{1985}]{jose85} Joseph R.~D., Wright G.~S., 1985, MNRAS, 214, 87 

\bibitem[\protect\citeauthoryear{Kaiser et al.}{2010}]{kais10} Kaiser N., Burgett W., Chambers K., Denneau L., Heasley J., Jedicke R., Magnier E., Morgan J., Onaka P., Tonry J., 2010, SPIE, 7733, 12 

\bibitem[\protect\citeauthoryear{Keel et al.}{1985}]{keel85} Keel W.~C., Kennicutt R.~C., Jr., Hummel E., van der Hulst J.~M., 1985, AJ, 90, 708 

\bibitem[\protect\citeauthoryear{Kelly \& Kirshner}{2011}]{kell11} Kelly P.~L., Kirshner R.~P., 2011, submitted, arXiv: 1110.1377v2

\bibitem[\protect\citeauthoryear{Kennicutt \& Keel}{1984}]{kenn84} Kennicutt R.~C., Jr., Keel W.~C., 1984, ApJ, 279, L5 

\bibitem[\protect\citeauthoryear{Kennicutt et al.}{1987}]{kenn87} Kennicutt R.~C., Jr., Roettiger K.~A., Keel W.~C., van der Hulst J.~M., Hummel E., 1987, AJ, 93, 1011 

\bibitem[\protect\citeauthoryear{Kewley, Geller \& Barton}{2006}]{kewl06} Kewley L.~J., Geller M.~J., Barton E.~J., 2006a, AJ, 131, 2004 

\bibitem[\protect\citeauthoryear{Kewley et al.}{2010}]{kewl10} Kewley L.~J., Rupke D., Zahid H.~J., Geller M.~J., Barton E.~J., 2010, ApJL, 721, 48 

\bibitem[\protect\citeauthoryear{Kiewe et al.}{2012}]{kiew12} Kiewe M., Gal-Yam A., Arcavi I., Leonard D.~C., Emilio Enriquez J., Cenko S.~B., Fox D.~B., Moon D.-S., Sand D.~J., Soderberg A.~M., CCCP T., 2012, ApJ, 744, 10

\bibitem[\protect\citeauthoryear{Klessen, Spaans, \& Jappsen}{2007}]{kles07} Klessen R.~S., Spaans M., Jappsen A.-K., 2007, MNRAS, 374, L29 

\bibitem[\protect\citeauthoryear{Kruijssen et al.}{2012}]{krui12} Kruijssen J.~M.~D., Pelupessy F.~I., Lamers H.~J.~G.~L.~M., Portegies Zwart S.~F., Bastian N., Icke V., 2012, MNRAS, 421, 1927

\bibitem[\protect\citeauthoryear{Kudritzki \& Puls}{2000}]{kudr00} Kudritzki R.-P., Puls J., 2000, ARA\&A, 38, 613

\bibitem[\protect\citeauthoryear{Larson \& Tinsley}{1978}]{lars78} Larson R.~B., Tinsley B.~M., 1978, ApJ, 219, 46 

\bibitem[\protect\citeauthoryear{Law et al.}{2009}]{law09} Law N.~M., Kulkarni S.~R., Dekany R.~G., Ofek E.~O., Quimby R.~M., Nugent  P.~E., Surace J., Grillmair C.~C., Bloom J.~S., Kasliwal M.~M.,  Bildsten L., Brown T., Cenko S.~B., Ciardi D., Croner E., Djorgovski S.~G., van Eyken J., Filippenko A.~V., Fox D.~B., Gal-Yam A., Hale D., Hamam N., Helou G., Henning J., Howell D.~A., Jacobsen J., Laher R., Mattingly S., McKenna D., Pickles A., Poznanski D., Rahmer G., Rau A., Rosing W., Shara M., Smith R., Starr D., Sullivan M., Velur V., Walters R., \& Zolkower J., 2009, PASP, 121, 1395

\bibitem[\protect\citeauthoryear{Leloudas et al.}{2011}]{lelo11} Leloudas G., Gallazzi A., Sollerman J., Stritzinger M.~D., Fynbo J.~P.~U., Hjorth J., Malesani D., Micha{\l}owski M.~J., Milvang-Jensen B., Smith M., 2011, A\&A, 530, A95 

\bibitem[\protect\citeauthoryear{Lennarz, Altmann \& Wiebusch}{2012}]{lenn12} Lennarz D., Altmann D., Wiebusch C., 2012, A\&A, 538, 120

\bibitem[\protect\citeauthoryear{Li, Mac Low \& Klessen}{2004}]{li04} Li Y., Mac Low M.-M., Klessen R.~S., 2004, ApJL, 614, 29

\bibitem[\protect\citeauthoryear{Li et al.}{2005}]{li05} Li W., Van Dyk S.~D., Filippenko A.~V., Cuillandre J.-C., 2005, PASP, 117, 121 

\bibitem[\protect\citeauthoryear{Li et al.}{2011}]{li11} Li W., et al., 2011, MNRAS, 412, 1441

\bibitem[\protect\citeauthoryear{Lotz, Primack \& Madau }{2004}]{lotz04} Lotz J.~M., Primack J., Madau P., 2004, AJ, 128, 168

\bibitem[\protect\citeauthoryear{Ma{\'{\i}}z-Apell{\'a}niz et al.}{2004}]{maiz04} Ma{\'{\i}}z-Apell{\'a}niz J., Bond H.~E., Siegel M.~H., Lipkin Y., Maoz D., Ofek E.~O., Poznanski D., 2004, ApJ, 615, L113 

\bibitem[\protect\citeauthoryear{Maund et al.}{2011}]{maun11} Maund J.~R., Fraser M., Ergon M., Pastorello A., Smartt S.~J., Sollerman J., Benetti S., Botticella M.~T., Bufano F., Danziger I.~J., Kotak R., Magill L., Stephens A.~W., Valenti S., 2011, ApJL, 739, L37 

\bibitem[\protect\citeauthoryear{Maund \& Smartt}{2005}]{maun05} Maund J.~R., Smartt S.~J., 2005, MNRAS, 360, 288 

\bibitem[\protect\citeauthoryear{Meurer et al.}{1995}]{meur95} Meurer G.~R., Heckman T.~M., Leitherer C., Kinney A., Robert C., Garnett D.~R., 1995, AJ, 110, 2665

\bibitem[\protect\citeauthoryear{Meynet et al.}{1994}]{meyn94} Meynet G., Maeder A., Schaller G., Schaerer D., Charbonnel C., 1994, A\&AS, 103, 97 

\bibitem[\protect\citeauthoryear{Meynet \& Maeder}{2003}]{meyn03} Meynet G., Maeder A., 2003, A\&A, 404, 975

\bibitem[\protect\citeauthoryear{Meynet \& Maeder}{2005}]{meyn05} Meynet G., Maeder A., 2005, A\&A, 429, 581

\bibitem[\protect\citeauthoryear{Mihos \& Hernquist}{1994}]{miho94} Mihos J.~C., Hernquist L., 1994, ApJ, 425, L13 

\bibitem[\protect\citeauthoryear{Mihos \& Hernquist}{1996}]{miho96} Mihos J.~C., Hernquist L., 1996, ApJ, 464, 641 

\bibitem[\protect\citeauthoryear{Milone et al.}{2008}]{milo08} Milone A.~P., Piotto G., Bedin L.~R., Sarajedini A., 2008, Mem.~Soc.~Astron.~Ital., 79, 623

\bibitem[\protect\citeauthoryear{Minkowski}{1941}]{mink41} Minkowski R., 1941, PASP, 53, 224 

\bibitem[\protect\citeauthoryear{Miralles-Caballero et al.}{2011}]{mira11} Miralles-Caballero D., Colina L., Arribas S., Duc P.~A., 2011, AJ, 142, 79

\bibitem[\protect\citeauthoryear{Modjaz et al.}{2011}]{modj11} Modjaz M., Kewley L., Bloom J.~S., Filippenko A.~V., Perley D., Silverman J.~M., 2011, ApJ, 731, L4 

\bibitem[\protect\citeauthoryear{Mokiem et al.}{2007}]{moki07} Mokiem M.~R., et al., 2007, A\&A, 473, 603

\bibitem[\protect\citeauthoryear{Muller et al.}{1992}]{mull92} Muller R.~A., Newberg H.~J.~M., Pennypacker C.~R., Perlmutter S., Sasseen T.~P., Smith C.~K., 1992, ApJ, 384, L9

\bibitem[\protect\citeauthoryear{Neff, Ulvestad \& Teng}{2004}]{neff04} Neff S.~G., Ulvestad J.~S., Teng S.~H., 2004, ApJ, 611, 186

\bibitem[\protect\citeauthoryear{Pastorello et al.}{2004}]{past04} Pastorello A., Zampieri L., Turatto M., Cappellaro E., Meikle W.~P.~S., Benetti S., Branch D., Baron E., Patat F., Armstrong M., Altavilla G., Salvo M., Riello M., 2004, MNRAS, 347, 74

\bibitem[\protect\citeauthoryear{Perets et al.}{2010}]{pere10} Perets H.~B., Gal-Yam A., Mazzali P.~A., Arnett D., Kagan D., Filippenko A.~V., Li W., Arcavi I., Cenko S.~B., Fox D.~B., Leonard D.~C., Moon D.-S., Sand D.~J., Soderberg A.~M., Anderson J.~P., James P.~A., Foley R.~J., Ganeshalingam M., Ofek E.~O., Bildsten L., Nelemans G., Shen K.~J., Weinberg N.~N., Metzger B.~D., Piro A.~L., Quataert E., Kiewe M., Poznanski D., 2010, Nature, 465, 322 

\bibitem[P{\'e}rez-Torres et al.(2009)]{pere09} P{\`e}rez-Torres, M.~A., Romero-Ca{\~n}izales, C., Alberdi, A., Polatidis A., 2009, A\&A, 507, 17 

\bibitem[\protect\citeauthoryear{Petrosian et al.}{2005}]{petr05} Petrosian A., Navasardyan H., Cappellaro E., McLean B., Allen R., Panagia N., Leitherer C., MacKenty J., Turatto M., 2005, AJ, 129, 1369 

\bibitem[\protect\citeauthoryear{Pettini \& Pagel}{2004}]{pett04} Pettini M., Pagel B.~E.~J., 2004, MNRAS, 348, L59 

\bibitem[\protect\citeauthoryear{Podsiadlowski, Joss, \& Hsu}{1992}]{pods92} Podsiadlowski P., Joss P.~C., Hsu J.~J.~L., 1992, ApJ, 391, 246 

\bibitem[\protect\citeauthoryear{Prantzos \& Boissier}{2003}]{pran03} Prantzos N., Boissier S., 2003, A\&A, 406, 259

\bibitem[\protect\citeauthoryear{Puls et al.}{1996}]{puls96} Puls J., Kudritzki R.~P., Herrero A., Pauldrach A.~W.~A., Hser S.~M., Lennon D.~J., Gabler R., Voels S.~A., Vilchez J.~M., Wachter S., Feldmeier A., 1996, A\&A, 305, 171 

\bibitem[\protect\citeauthoryear{Rampazzo et al.}{2005}]{ramp05} Rampazzo R., Plana H., Amram P., Bagarotto S., Boulesteix J., Rosado M., 2005, MNRAS, 356, 1177

\bibitem[\protect\citeauthoryear{Rau et al.}{2009}]{rau09} Rau A., Kulkarni S.~R., Law N.~M., Bloom J.~S., Ciardi, D., Djorgovski G.~S., Fox D.~B., Gal-Yam A., Grillmair C.~C., Kasliwal M.~M., Nugent P.~E., Ofek E.~O., Quimby R.~M., Reach W.~T., Shara M., Bildsten L., Cenko S.~B., Drake A.~J., Filippenko A.~V., Helfand D.~J., Helou G., Howell D.~A., Poznanski D.\& Sullivan M.  2009, PASP, 121, 1334

\bibitem[\protect\citeauthoryear{Rich et al.}{2012}]{rich12} Rich J.~A., Torrey P., Kewley L.~J., Dopita M.~A., Rupke D.~S.~N., 2012, arXiv:1204.5520

\bibitem[\protect\citeauthoryear{Rupke, Kewley \& Chien}{2010}]{rupk10} Rupke D.~S.~N., Kewley L.~J., Chien L.-H., 2010, ApJ, 723, 1255

\bibitem[\protect\citeauthoryear{Sanders \& Mirabel}{1996}]{sand96} Sanders D.~B., Mirabel I.~F., 1996, ARA\&A, 34, 749 

\bibitem[\protect\citeauthoryear{Schmidt}{1959}]{schm59} Schmidt M., 1959, ApJ, 129, 243 

\bibitem[\protect\citeauthoryear{Shaw}{1979}]{shaw79} Shaw R.~L., 1979, A\&A, 76, 188 

\bibitem[\protect\citeauthoryear{Smartt}{2009}]{smar09} Smartt S.~J., 2009, ARA\&A, 47, 63 

\bibitem[\protect\citeauthoryear{Smartt et al.}{2009}]{smart09} Smartt S.~J., Eldridge J.~J., Crockett R.~M., Maund J.~R., 2009, MNRAS, 395, 1409

\bibitem[\protect\citeauthoryear{Smith et al.}{2011}]{smit11} Smith N., Li W., Filippenko A.~V., Chornock R., 2011, MNRAS, 412, 1522

\bibitem[\protect\citeauthoryear{Sollima et al.}{2007}]{soll07} Sollima A., Beccari G., Ferraro F.~R., Fusi Pecci F., Sarajedini A., 2007, MNRAS, 380, 781

\bibitem[\protect\citeauthoryear{Sollima}{2008}]{soll08} Sollima A., 2008, MNRAS, 388, 307

\bibitem[\protect\citeauthoryear{Sollima et al.}{2010}]{soll10} Sollima A., Carballo-Bello J.~A., Beccari G., Ferraro F.~R., Pecci F.~F., Lanzoni B., 2010, MNRAS, 401, 577

\bibitem[\protect\citeauthoryear{Surace}{1998}]{sura98} Surace J.~A., 1998, Ph.D. thesis, Univ. Hawaii

\bibitem[\protect\citeauthoryear{Th{\"o}ne et al.}{2009}]{thon09} Th{\"o}ne C.~C., Micha{\l}owski M.~J., Leloudas G., Cox N.~L.~J., Fynbo J.~P.~U., Sollerman J., Hjorth J., Vreeswijk P.~M., 2009, ApJ, 698, 1307

\bibitem[\protect\citeauthoryear{Tsvetkov, Pavlyuk, \& Bartunov}{2004}]{tsve04} Tsvetkov D.~Y., Pavlyuk N.~N., Bartunov O.~S., 2004, Astronomy Letters, 30, 729

\bibitem[\protect\citeauthoryear{van den Bergh}{1997}]{vand97} van den Bergh S., 1997, AJ, 113, 197 

\bibitem[\protect\citeauthoryear{van den Bergh, Li \& Filippenko}{2005}]{vand05} van den Bergh S., Li W., Filippenko, A.~V., 2005, PASP, 117, 773 

\bibitem[\protect\citeauthoryear{Van Dyk}{1992}]{vanD92} Van Dyk S.~D., 1992, AJ, 103, 1788 

\bibitem[\protect\citeauthoryear{Van Dyk, Hamuy \& Filippenko}{1996}]{vanD96} Van Dyk S.~D., Hamuy M., Filippenko A.~V., 1996, AJ, 111, 2017 

\bibitem[\protect\citeauthoryear{Van Dyk et al.}{2000}]{vand00} Van Dyk S.~D., Peng C.~Y., King J.~Y., Filippenko A.~V., Treffers R.~R., Li W., Richmond M.~W., 2000, PASP, 112, 1532

\bibitem[\protect\citeauthoryear{Veilleux, Kim \& Sanders}{2002}]{veil02} Veilleux S., Kim D.-C., Sanders D.~B., 2002, ApJS, 143, 315

\bibitem[\protect\citeauthoryear{Yoon, Woosley \& Langer}{2010}]{yoon10} Yoon S.-C., Woosley S.~E., Langer N., 2010, ApJ, 725, 940 

\bibitem[\protect\citeauthoryear{Zamojski et al.}{2011}]{zamo11} Zamojski M., Yan L., Dasyra K., Sajina A., Surace J., Heckman T., Helou G., 2011, ApJ, 730, 125

\end{thebibliography}

\label{lastpage}

\end{document}